\documentclass[twocolumn,times,tighten]{aastex63}

\usepackage{color, enumitem, hyperref, verbatim, amsmath, amstext, mathrsfs, hypcap, afterpage, marginnote, makecell, subfigure}
\usepackage{multirow}
\usepackage{booktabs}
\usepackage{natbib}
\usepackage{comment}
\usepackage{graphicx}	% Including figure files
\usepackage{amsmath}	% Advanced maths commands
\usepackage{amssymb}	% Extra maths symbols
\usepackage{lineno}

\graphicspath{{./}{figures/}}

\received{\today}
\shorttitle{Sub-Alfvenic Turbulence and Magnetic Field Strength}
\shortauthors{Lazarian et al.}

\begin{document}

\title[Violation of Energy equi-partition in sub-Alfvenic incompressible MHD Turbulence]{Sub-Alfvenic Turbulence: Magnetic to Kinetic Energy Ratio, Modification of Weak Cascade \\and Implications for Magnetic Field Strength Measurements}

\author[0000-0002-7336-6674]{A. Lazarian}
\affiliation{Department of Astronomy, University of Wisconsin-Madison, Madison, WI, 53706, USA}
\email{lazarian@astro.wisc.edu}

\author[0000-0003-3328-6300]{Ka Wai Ho}
\affiliation{Department of Astronomy, University of Wisconsin-Madison, Madison, WI, 53706, USA}
\affiliation{Theoretical Division, Los Alamos National Laboratory, Los Alamos, NM 87545, USA}
\email{kho33@wisc.edu}

\author[0000-0003-1683-9153]{Ka Ho Yuen}
\affiliation{Theoretical Division, Los Alamos National Laboratory, Los Alamos, NM 87545, USA}
\email{kyuen@lanl.gov}

\author[0000-0002-2307-3857]{Ethan Vishniac}
\affiliation{Physics Department, Johns Hopkins University, USA}
\email{evishni1@jhu.edu}

% These dates will be filled out by the publisher
\date{Accepted XXX. Received YYY; in original form ZZZ}

% why there is no mnras bib
%\shorttitle{Magnetic-to-Kinetic Energy Ratio in sub-Alfvenic turbulenc cascade}

\begin{abstract}
We study the properties of sub-Alfvenic magnetohydrodynamic (MHD) turbulence, i.e., turbulence with Alfven Mach number $M_A=V_L/V_A<1$, where $V_L$ is the velocity at the injection scale and $V_A$ is the Alfven velocity.
We demonstrate that MHD turbulence can have different regimes depending on whether it is driven by velocity or magnetic fluctuations. Suppose the turbulence is driven by isotropic bulk forces, i.e., velocity-driven, in an incompressible conducting fluid. In that case, we predict that the kinetic energy is $M_A^{-2}$ times larger than the energy of magnetic fluctuations. This effect arises from the long parallel wavelength tail of the forcing, which excites modes with $k_\|/k_\perp < M_A$. We also predict that as the turbulent cascade reaches the strong regime, the energy of slow modes exceeds the energy of Alfven modes by a factor $M_A^{-1}$. These effects are absent if the turbulence is magnetically driven at the injection scale.  We confirm these predictions with numerical simulations. As the assumption of magnetic and kinetic energy equipartition is at the core of the Davis-Chandrasekhar-Fermi (DCF) approach to measuring magnetic field strength in sub-Alfvenic turbulence, we conclude that the DCF technique is not universally applicable. In particular, we suggest that the dynamical excitation of long azimuthal wavelength modes in the galactic disk may compromise the use of the DCF technique. We discuss alternative expressions that can be used to obtain magnetic field strength from observations.
\end{abstract}

\section{Introduction} 

Magnetized astrophysical fluids are turbulent due to the high Reynolds number of flows \citep{2004tise.book.....D}. The motions are described as plasma turbulence on small scales, i.e., ion inertia scale. At the same time, the Magneto Hydrodynamic (MHD) approach has been successfully applied to describe processes for describing astrophysical phenomena in a variety of astrophysical settings. The magnetic field constrains the motions of charged particles perpendicular to the direction of the magnetic field. It decreases the viscosity perpendicular to the magnetic field of both fully and partially ionized gas (see Table 1. in \cite{Eyink2011}). Observations testify that MHD turbulence is ubiquitous (see  \citealt{2010ASPC..429..113L,BurL21}) and affects key astrophysical processes, including the gas and phase transfer in dynamics of interstellar media (see \citealt{filament_review}), star formation (see \citealt{MO07}) and cosmic ray propagation (see \citealt{2023FrASS..1054760L}). 

In many astrophysical environments, MHD turbulence is sub-Alfvenic, i.e., Alfven velocity exceeds the turbulent velocity, making magnetic effects dominant for the turbulent motions. This subject has been studied less than transAlfvenic turbulence. This motivates our present study of sub-Alfvenic turbulence.  

Astrophysical turbulence happens in compressible media. However, the relative importance of compressibility effects differs from case to case. Many vital properties of MHD turbulence can be understood disregarding compressibility, i.e., considering {\it incompressible} MHD. This is similar to hydrodynamics, where Kolmogorov theory, i.e., {theory of {\it incompressible} turbulence}, successfully explains many basic phenomena. Note that the theory of compressible hydrodynamic turbulence stays a rather unsettled subject (see \cite{2004tise.book.....D}).  Thus, we focus on the properties of incompressible sub-Alfvenic turbulence for the present study. 

Incompressible turbulence is at the core of the MHD turbulence theory (see \cite{BL19}, for a monograph). The properties of the Alfvenic mode cascade are marginally affected by media compressibility \citep{CL02_PRL}. In a compressible medium, it slaves slow modes imposing its scaling on them (\citealt{GS95}, henceforth GS95, \citealt{LG01}, \citealt{CL02_PRL,CL03}). 
This makes the scaling of slow modes independent of compressibility effects. Alfvenic turbulence
also determine magnetic field wandering (\citealt{LV99}, LV99), giving rise to super-diffusive dispersion of magnetic field lines (LV99, \citealt{Eyink2011,Eyink13}). In particular, the magnetic field wandering determines the observed variations of the direction of the dust polarization employed for obtaining magnetic field strength using the technique proposed by \cite{1951PhRv...81..890D} and \cite{ChanFer53}. The corresponding technique is widely used to obtain magnetic field strength for sub-Alfvenic, i.e., with turbulent injection velocity $V_L$ less than Alfven velocity $V_A$ molecular clouds.  Further on, we refer to the aforementioned technique as the {\it DCF technique}.

In the DCF technique, the 2D magnetic field direction dispersion measured through dust polarimetry is combined with the velocity dispersion measured spectroscopically from the same molecular cloud. The technique obtains magnetic field strength assuming the turbulence is Alfvenic, isotropic and has an equipartition between the magnetic and kinetic energies. 

The equipartition of magnetic field fluctuations and velocity fluctuations has been challenged empirically based on the analysis of {\it compressible} sub-Alfvenic MHD turbulence simulations \citep{Federrath_2016,2022MNRAS.515.5267B}. These and further studies attributed the energy disparity and other non-classical properties of observed turbulence to the effects of {\it compressibility}. This resulted in a suggestion that the traditional DCF technique is valid for incompressible Goldreich-Sridhar turbulence, while the technique should be modified to deal {\it compressible MHD turbulence} in molecular clouds \citep{2021A&A...647A.186S, 2021A&A...656A.118S}.  The theoretical arguments supporting this modified technique require turbulence to be compressible.\footnote{Effects of compressibility are complicated and were frequently appealed to when researchers faced unexpected phenomena. For instance, the numerical discovery of MHD turbulence fast decay was surprising and originally attributed to energy transfer from Alfvenic to compressible dissipative motions (see \cite{Stone98}). In reality, this was a fundamental property of strong MHD turbulence, which is also present in the incompressible limit (see \cite{CL02_PRL}).}

 In this paper, we do not consider fluid compressibility and thus, we do not discuss the theories that appeal to compressibility to either describe the aforementioned properties of turbulence or justify the "compressible" modification of the DCF technique. Instead, we consider {\it incompressible} MHD turbulence and explore how its properties are affected by velocity versus magnetic driving at the injection scale.  We test our theoretical finding by analyzing numerical simulations of {\it incompressible} 3D MHD turbulence. 

Our study identifies the velocity driving of MHD turbulence as the source of violating kinetic and magnetic turbulent energies equipartition. This effect was not reported earlier in incompressible MHD, as in most of the corresponding theory-motivated numerical studies of incompressible MHD turbulence (see \citealt{2012MNRAS.422.3495B,Bere14}), 
%https://ui.adsabs.harvard.edu/abs/2012MNRAS.422.3495B/abstract
%https://ui.adsabs.harvard.edu/abs/2014ApJ...784L..20B/abstract
the turbulence is driven via Elsasser variables, where magnetic field fluctuations are induced at the injection scale. 

A good correspondence of some of our results to the ones identified earlier as arising from the compressibility of MHD turbulence is suggestive that the fluid compressibility may not be the reason for the reported "non-classical" behavior of MHD turbulence. As astrophysical simulations of turbulence are made with compressible codes  that employed velocity driving \citep{2008ApJ...688L..79FZ, Federrath_2016, 2021NatAs...5..365F,2020MNRAS.498.1593B,2022MNRAS.515.5267B},
 we conjecture that the reported effects can universally arise from how turbulence is driven.

In what follows, in \S \ref{sec:subAlfven}, we describe how the properties of sub-Alfvenic turbulence near the injection scale can be modified by velocity driving. We describe our numerical simulations in \S \ref{sec:num} and test our theoretical prediction with 3D MHD numerical simulations of incompressible turbulence in \S \ref{sec:analysis}. In \S \ref{sec:comp} and \S \ref{sec:magfluc}, we compare our findings with the results of earlier numerical simulations. The implications of our results for obtaining magnetic field strength from observations are discussed in \S \ref{sec:imp}. We discuss our results in \S \ref{sec:disc} with the summary in \S \ref{sec:sum}. 

\section{Sub-Alfvenic MHD turbulence}
\label{sec:subAlfven}

Kolmogorov theory is the theory of hydrodynamic {\it incompressible} turbulence, which showed a lot of practical applicability in spite of its assumption of incompressibility. Below we discuss the strong {\it incompressible} Alfvenic turbulence case to show its analogy the Kolmogorov theory. This will provide intuitive guidelines for addressing the case of weak Alfvenic turbulence that demonstrates more facets depending on its driving and boundary conditions.

\subsection{Lessons from strong Alfvenic turbulence}

\subsubsection{Local system of reference for turbulent motions}

The theory of MHD turbulence can be traced to the classical works studied by \cite{Iro64} and \cite{Kra65}, which, however, missed the major of its property, i.e., its anisotropy (see \citealt{Hig84} ). In GS95, the theory of MHD turbulence describing its anisotropic nature was formulated for Alfven Mach number 
\begin{equation}
    M_{A,b}\equiv \frac{V_{L,b}}{V_A}=\frac{\delta B_L}{B},
    \label{MA,b}
\end{equation}
equal 1, where  $V_{L,b}$ is the injection velocity of Alfvenic or pseudo-Alfvenic fluctuations at the injection scale $L$, and $V_A$ is the Alfven velocity. Pseudo-Alfvenic fluctuations are the incompressible limit of slow MHD fluctuations, and they have fluctuations in velocity and magnetic field related through the Alfven relation. Thus, the second relation in Eq. (\ref{MA,b}) expressing $M_{A,b}$ follows automatically. In other words, GS95 is the theory of incompressible MHD turbulence in a trans-Alfvenic regime.  

The original formulation's frequently overlooked limitation is that it did not include a fundamental notion of the {\it local} system of reference. The closure relations justifying the GS95 relations were obtained in the generally accepted reference frame related to the mean magnetic field at that time. The problem is that, in reality, the fluctuations parallel and perpendicular to the mean magnetic field do not follow to GS95 scaling, as was shown in subsequent numerical studies (see \citealt{CV00}). Instead, magnetic fluctuations must be measured in the {\it local} system of reference defined by the direction of the magnetic field in the direct vicinity of the perturbation. The necessity of measuring turbulent fluctuations with respect to the direction of the magnetic field that directly affects these fluctuations became obvious in LV99 where the inseparable connection of Alfvenic turbulence and turbulent reconnection was demonstrated. In numerical studies by  \citep{CV00,MG01}, it was demonstrated that it is only in the local system of reference that the GS95 relation between the parallel and perpendicular to magnetic field turbulent motions is satisfied. In what follows, we denote these scales $l_\|$ and $l_\bot$, keeping in mind that they are measured with respect to the local direction of the magnetic field.

\subsubsection{Strong MHD turbulence for $M_{A,b}<1$}

The incompressible MHD turbulence theory extension for the regime of $M_{A,b}<1$ in LV99 was performed in LV99. In the latter study, it was shown that Alfvenic turbulence with $M_{A,b}<1$ exhibits at the injection scale the {\it weak} regime, which transfers to the {\it strong} GS95-like regime for smaller scales. The "weak" and "strong" are understood in terms of non-linear interactions, i.e., the relation of the cascading time $\tau_{cas}$, and the period of Alfvenic perturbation $t_A\sim \omega^{-1}\sim l_\|/V_A$. The weak turbulence corresponds to $\tau_{cas}>t_A$, and the strong turbulence corresponds to $\tau_{cas}\sim t_A$. The two regimes are not related to fluctuations' amplitude, and turbulence with a large amplitude of fluctuations can be in the {\it weak} regime, while turbulence with a small amplitude can be in the {\it strong} regime. In what follows, we present an intuitive picture of the strong MHD turbulence cascade from the LV99 theory of turbulent reconnection (see \citealt{Laz20} for a review). 

Kolmogorov hydrodynamic turbulence \citep{Kol41} can be described as eddies following Richardson's intuitive picture (see \citealt{richardson1922weather}). According to this picture, the large eddies generate smaller eddies until the viscosity damps the smallest eddies.  Transferring energy from the driver to eddies is advantageous due to minimal resistance to the induced motions. On the contrary, inducing wave motions requires actions against the restoring force. Thus, given a choice, the kinetic energy is expected to transfer to eddies if the eddy formation is possible. 

In the presence of the magnetic field, it was traditionally assumed that the field prevented eddy motions. Thus, the case of MHD turbulence was traditionally considered in terms of interacting waves \citep{Iro64, Kra65}. This, however, disregarded the possibility of exciting the eddy-like motions mixing magnetized fluid perpendicular to the direction of the magnetic field. Such eddies' existence follows the turbulent reconnection theory (LV99). 

For turbulent fluid, the direction of the magnetic field varies in space. Therefore, to define the direction of these mixing motions, one should consider the magnetic field in the eddy vicinity, i.e., to define the {\it local} direction of the magnetic field. This revealed an analogy between the hydrodynamic and Alfvenic turbulence, making evident why, in GS95, the Kolmogorov spectrum of velocity fluctuations is expected.
\footnote{The discussion whether the turbulence has Iroshnikov-Kraichnan $\sim k^{-3/2}$ or Kolmogorov $\sim k^{-5/3}$ spectrum proceeded for several years (see \citep{Boldyrev,2007APS..DPPPO7001B,2006ApJ...640L.175B,BL09}, but in our opinion, the compelling evidence for the Kolmogorov spectrum is demonstrated in \cite{Bere14}. Note that the theories advocating the Iroshnikov-Kraichnan spectrum predict anisotropies that strongly disagree with numerical simulations \citep{BL19}.}
%Richardson L.F. 1922, Weather prediction by Numerical process", Cambridge University Press

If one considers Alfvenic eddy-type motion of scale $l_\bot$ mixing magnetic field perpendicular to the direction of the local magnetic field, the kinetic energy of turbulent motions at the scale $l_\bot$ is $\sim \rho v_l^2$, where $\rho$ is constant for incompressible turbulence. In terms of hydrodynamic equations, the non-linear term for perpendicular motions that we consider is $(v\nabla)v\sim v_\bot^2/l_\bot$ which can be rewritten as $v_\bot\omega_{nl}$, where $\omega_{nl}\sim v_\bot/l_\bot$. 

It is natural to associate $\tau_{cas}$ with the non-linear time of the eddy turnover time $\tau_{turn}\sim l_\bot/v_l$. The latter condition is satisfied in the presence of turbulent reconnection (LV99) that predicts the reconnection time of the magnetic field in the eddy to be the eddy turnover time $\tau_{turn}$ for the eddies rotating perpendicular to the direction of the magnetic field. Thus, $v_l$ should be identified with the eddy rotation perpendicular to the magnetic field. The hydrodynamic-like cascading of energy involving such eddies corresponds to $v_l^2/\tau_{cas}\sim v_l^3/l =const$, providing the well-known Kolmogorov scaling of velocity fluctuations $v_l\sim l^{1/3}.$

Overall, the cascading of eddies aligned with the ambient magnetic field is similar to that of hydrodynamic eddies. Thus, the hydrodynamic-like character is relevant to perpendicular to magnetic field motions. The direction of the magnetic field surrounding the eddy, i.e., the local magnetic field.  The latter can be practically defined using structure functions \citep{CV00}. 

\subsubsection{Scaling and anisotropy for strong MHD turbulence for $M_{A,b}<1$}
\label{scal}

The eddy motions of the magnetized fluid induce Alfven waves with the period $\sim l_\|/V_A$, where $l_\|$ is the parallel scale of the eddy, which, as we mentioned earlier, should be measured in respect to the {\it local} direction of the magnetic field. Therefore, the {\it critical balance} condition postulated in GS95, should be written in the local system of reference as
\begin{equation}
    l_\bot/V_l\approx l_\|/V_A.
    \label{balance}
\end{equation} 
Eq. (\ref{balance}) reflects that the magnetic field eddy mixing perpendicular to the magnetic field induces the Alfvenic perturbation that defines the parallel to the magnetic field eddy scale. Recalling the definitions of the non-linear and Alfvenic interactions, Eq. (\ref{balance}) can be rewritten as $\omega_{nl}\approx t_A^{-1}$.

Naturally, the condition given by Eq. (\ref{balance}) is not satisfied if the motions are considered in the mean magnetic field frame. Incidentally, in the eddy picture, the decay of {\it strong} Alfvenic turbulence the decay in one Alfven crossing time is expected, as it is well-known that hydrodynamic turbulence decays over one eddy turnover time.
%https://ui.adsabs.harvard.edu/abs/1998ApJ...508L..99S/abstract

To summarize, the strong Alfvenic turbulence, e.g., the turbulence corresponding to $M_{A,b}=1$, as the case for the GS95 theory, can be presented as a collection of eddies aligned with the magnetic field surrounding the eddies. Thus, the eddies aligned with the surrounding magnetic field are involved in hydrodynamic-type motion. Naturally, the energy is channeled along this path of minimal resistance, and thus, the energy of random driving mostly resides with the eddy cascade.

Consider a case of isotropic turbulence driving for strongly magnetized medium, i.e., for $M_{A,b}<1$. It is obvious that if both injection parallel and perpendicular scales are equal to $L$, the critical balance condition given by Eq. (\ref{balance}) cannot be satisfied. Thus, the GS95 theory of strong Alfvenic turbulence is not applicable at the injection scale. This is the domain of weak Alfvenic turbulence that we consider further in \S \ref{sec:weakturb}.

At the same time, there are no limitations to applying the critical balance to sufficiently small eddies. 
As the turbulent injection velocities decrease compared to Alfven velocity $V_A$, the eddies get more elongated parallel to the local magnetic field direction.  LV99 showed that in the latter case, the GS95 relations are modified to include Alfven Mach number dependence:
\begin{equation}
    v_l\approx V_L \left(\frac{l_\bot}{L}\right)^{1/3} M_{A,b}^{1/3},
    \label{vl}
\end{equation}
and 
\begin{equation}
    l_\|\approx L \left(\frac{l_\bot}{L}\right)^{2/3} M_{A,b}^{-4/3},
    \label{lpar}
\end{equation}

An essential property of the cascade is that according to Eq. (\ref{vl}), the rate of non-linear interactions increases with the decrease of the scale, i.e., $\omega_{nl}\sim l_\bot^{-2/3}$. This means this strength increases with the decrease of the scale of the turbulent motions, even though the amplitude of the fluctuations decreases.

We are interested in our present study's $M_{A,b}<1$ regime. First, it is important to define the applicability of the critical balance. It is evident from Eq. (\ref{lpar}) that for perpendicular scale 
\begin{equation}
l_{tr} = LM_{A,b}^2,
\label{ttr}
\end{equation}
the parallel scale of magnetic eddies reaches the injection scale $L$. This value of $l_\bot$ defines the boundary of the applicability of the magnetic eddy type turbulence, or the {\it strong MHD turbulence}. Note that the perpendicular cascade corresponds to the Kolmogorov one in terms of the energy spectrum. At the same time, it is easy to see that in the strong cascade $v_l$ scales as $l_\|^{1/2}$ in terms of parallel to magnetic field perturbations, which corresponds to the parallel to magnetic field spectrum $\sim k_\|^{-2}$. When the spectra are averaged for a given $|k|$, the dominant contribution comes from the perpendicular fluctuations
providing $E_{strong} (|k|) \sim k^{-5/3}$.

At the scale $l_{tr, \bot}$ for $M_{A.b}<1$, the velocity value is $V_L M_{A,b}$, i.e., smaller than the injection velocity by a factor $M_{A,b}<1$. The fluctuations in the strong turbulence regime are Alfvenic with magnetic field and velocity connected by the Alfvenic relation, i.e., equipartition is naturally satisfied. Thus, the amplitude of fluctuation of the magnetic field at $l_{tr, \bot}$ is
\begin{equation}
    \delta B_{max, strong}\approx \frac{V_L}{\sqrt{4 \pi \rho}} M_{A,b}.
\label{max_strong}
\end{equation}
The picture changes for scaling at scales larger than $l_{tr}$. This is a regime of {\it weak turbulence} (see LV99, \citealt{2000JPlPh..63..447G,2011LNP...825.....N}). The parallel scale of Alfvenic waves induced by driving at the injection scale $L$ exceeds $L$ at these scales. As we discuss further, this requirement makes the weak turbulence less robust than the strong one. 

GS95 demonstrated that pseudo-Alfven modes are passive and are slaved by Alfvenic cascade. Thus, the relations between the parallel and perpendicular scales of pseudo-Alfven fluctuations are given by Eq. (\ref{balance}). This conclusion is also true for $M_{A,b}<1$ case for $l_\bot<l_{tr}$. 

\subsection{Weak turbulence driven by velocities: the violation of the energy equi-partition}
\label{sec:weakturb}

\subsubsection{Stiff magnetic field lines}
\label{2.2.1}

Energy can be injected into MHD turbulence in different forms. Below, we focus on the effects of energy injection on velocity fluctuations. 

As $M_{A,b}$ decreases, the kinetic motions gradually have more difficulty in magnetic field bending on the injection scale. Consider an extreme case of velocity driving corresponding to $M_{A,b}\rightarrow 0$ limit {\it assuming magnetic field lines are stiff}. In other words, the timescale for perturbing magnetic field lines, parametrized by $\tau=(k_\parallel v_A)^{-1}$, is very large.  In this setting, the only possible motions of magnetized conducting fluid correspond to the mixing of absolutely straight magnetic field lines, i.e., the motions with the parallel wavenumber $k_\|=0$. Such {\it perpendicular} motions do not induce Alfven waves or any other finite-time magnetic field fluctuations. No magnetic field bending corresponds to no energy of Alfven waves. Within our thought experiment, in the limit of $k_\parallel=0$, turbulent magnetic energy vanishes, in contrast with the case of strong turbulence for which magnetic $E_B$ and kinetic $E_K$ energies are in equipartition. In what follows, we consider finite $M_{A,b}<1$.  The magnetic field lines, in this case, are not rigid, and magnetic turbulent energy is not zero. Nevertheless, through continuity, our thought experiment suggests that the equipartition sub-Alfvenic turbulence is not guaranteed, and for some types of driving, $E_K>E_B$ is expected.

{To elucidate the basic properties of subAlfvenic MHD turbulence, our study deals with the incompressible case. Nevertheless, magnetic field compressions are possible in this case. In the thought experiment of an extremely rigid magnetic field above, one can consider nearly straight magnetic field lines being compressed perpendicularly. These bunches of magnetic field lines present the $k_\|=0$ incarnation of pseudo-Alfven/slow modes. Unlike its Alfvenic counterpart for $M_A\rightarrow 0$, these motions are associated with turbulent magnetic energy. Thus, one can expect that slow modes' magnetic energy can exceed Alfvenic modes' energy for subAlfvenic turbulence. }

Another effect not considered in the theory of strong turbulence is the fluid motion along the magnetic field lines. In the limit of $M_{A,b}\rightarrow 0$ and a random isotropic velocity driving at the injection scale $L$, this driving excites only {\it parallel} kinetic motions without any magnetic field perturbations. This effect persists for $M_{A,b}<1$ as the random isotropic driving tends to move fluid in the direction of the least resistance, the direction along the magnetic field lines. The effect disappears for $M_{A,b}> 1$ as the magnetic field gets pliable to bending.
Thus, random velocity driving for small $M_{A,b}$ can induce $E_K\gg E_B$ in terms of both parallel and perpendicular to magnetic field fluctuations. This effect is unrelated to fluid compressibility, so in this study, we do not discuss the theories that attribute the numerically detected disparity of kinetic and magnetic energies to fluid compressibility (see \citealt{2022MNRAS.515.5267B, 2023A&A...673A..76S}).

One can consider magnetic field lines as parallel, nearly rigid tubes to understand driving low $M_{A,b}$ turbulence. Such intuitive treatment is congruent with several recent in-situ solar wind studies \citep{2013PhRvE..87a3019B,2019MNRAS.488..859Y,2024NatAs.tmp...71Z,2024ApJ...962...89Z}. The perpendicular motion of these tubes marginally deforms them. At the same time, the fluid can be pumped along the tubes without deforming them much, either. Both types of motion involve kinetic motions while inducing a marginal energy transfer into the elastic stress energy. The latter is the analog of the magnetic perturbation energy.

The fluid flow along magnetic field lines  causes variations in the magnetic field tension.  For a flow with velocity $V_{L,\|}$ along magnetic field lines, the Bernoulli equation (see \citealt{2000ifd..book.....B}) in the absence of gravity provides $P+\rho V^2/2 = const$, where $P$ is the pressure of the fluid. {If the velocity of fluid moving along the magnetic tube embedded in the media with external pressure $P_0$ changes in point 1 from $V_1$ to $V_2$ at point 2, the change of the fluid pressure changes from $P_1$ to $P_2$ according to the equation:
\begin{equation}
    P_0 + P_1 +1/2 \rho V_1^2 = P_0+ P_2 +1/2\rho V_2,
    \label{Bern1}
\end{equation}
resulting in 
\begin{equation}
    B\delta B \approx 1/2\rho (V_2^2-V_1^2),
    \label{Bern2}
\end{equation}
where we accounted that the changes in the fluid pressure induce the changes in the magnetic pressure according to $P_1-P_2=\delta P\approx B\delta B$. The fluctuation $\delta B$ can be positive or negative depending on whether the amplitude of $V_1$ or $V_2$ is larger. Taking squares of both parts of Eq. (\ref{Bern2}), one can determine the variance of the magnetic field fluctuations
$(\delta B)^2$ and the standard deviation, which is the square root of the variance, i.e., $|\delta B|$.

For turbulent driving, it is natural to associate the maximal difference in velocities with the velocity at the driving scale, i.e., $V_{L,\|}^2\approx max\{V_2^2-V_1^2\}$. Therefore one can write the resulting equation as

}
\begin{equation}
    \rho V_{L, \|}^2 \approx B |\delta B_\||.
    \label{2}
\end{equation}
 The regions corresponding to higher fluid pressure result in the bulging of magnetic field lines, locally decreasing $\delta B_\|$, and the regions of lower fluid pressure result in getting magnetic field lines closer together, increasing magnetic field back-reaction. 

Later, in \S \ref{sec:strange}, we discuss that while Eq. (\ref{2}) resembles Eq. (\ref{strange}), which was used in \cite{2021A&A...647A.186S}, their physical reasoning and justification are very different from ours. In our approach, Eq. (\ref{2}) describes the response of the magnetic field to the change of the pressure in the flow of fluid that is independent of turbulence compressibility. At the same time, in 
\cite{2021A&A...647A.186S}, Eq. (\ref{strange}) is employed to describe the balance of kinetic and magnetic energies in a {\it compressible} MHD turbulence. In \S \ref{sec:strange}, we acknowledge our confusion with the theoretical arguments justifying the latter approach. At the same time, our derivation of Eq. (\ref{2}) also holds in the compressible case, which may reflect the universal property of sub-Alfvenic turbulence driven by isotropic velocity fluctuations. 

The fluctuations of the magnetic field 
 parallel to the mean magnetic field $\delta B_\|$ induce the magnetic field fluctuations perpendicular to the mean field direction $\delta B_\bot$. This is the consequence of the divergence-free magnetic field, i.e., $\nabla \delta { B=0}$. Taking the z-axis along the mean magnetic field ${ B}$, one can write the divergence in the Cartesian coordinate system:
\begin{equation}
    \frac{\partial \delta B_x}{\partial x}+\frac{\partial \delta B_y}{\partial y}+\frac{\partial \delta B_z}{\partial z}=0,
\end{equation}
where $\delta B_\bot=\sqrt{\delta B_x^2+\delta B_y^2}$. At the injection scale, one can write $\partial \delta B_i\approx \delta B_i$, and $\partial x_i\approx L$, where $i=x,y,z$, where $L$ is the scale of isotropic turbulence driving. This results in $-\delta B_x = \delta B_y + \delta B_z$ at the injection scale. Substituting this in the expression of perpendicular fluctuation, one gets for the energy of perpendicular fluctuations $\delta B_\bot^2=2\delta B_y^2 +2\delta B_z \delta B_y +\delta B_z^2$. This change in the energy is a magnetic field response to a perturbation $\delta B_\|$. Naturally, the magnetic field resists perturbations and minimizes the increase of the energy induced by $\delta B_\|$.\footnote{Formally, the divergence-free condition can also be satisfied if $\delta B_\bot$ is much larger than $\delta B_\|$. However, in the setting when hydrodynamic flows induce $\delta B_\|$ and $\delta B_\bot$ is induced to ensure the magnetic field is divergence-free, the solutions with $\delta B_\bot$ larger than $\delta B_\|$ are not physical.}
For a given $\delta B_z$, the minimum of this energy corresponds to $\delta B_y\approx -\delta B_z/2$. The corresponding value of $\delta B_x \approx -\delta B_z/2$, which results in 
\begin{equation}
\delta B_\bot \approx \delta B_\|.
\label{3}
\end{equation}

\begin{comment}
In the case of isotropic driving, $V_{L,\bot}\approx V_{L,\|}$ which entails $\delta B_\bot\approx \delta B_\|$, i.e., the amplitudes of large scale perpendicular and parallel magnetic field fluctuations are similar.

Both Eq. (\ref{1}) and Eq. (\ref{2}) are not valid if magnetic perturbations drive the turbulence. In the latter case, the velocities and magnetic field fluctuations have an Alfvenic nature with $\rho V_{L,\bot}\sim \delta B_{\bot}$ and $\rho V_{L,\|}\sim \delta B_\|$. Thus, proves that the properties of sub-Alfvenic magnetic turbulence depend on whether the magnetic or velocity fluctuations are driven at the injection scale. Eq. (\ref{1}) and Eq.(\ref{2}) suggest that the turbulent magnetic energy for sub-Alfvenic turbulence driven by velocity fluctuations is a factor $\sim M_{A,b}$ less the turbulent kinetic energy. Further in the paper, we discuss in detail the relations between different cascades induced in velocity-driven sub-Alfvenic turbulence.
\end{comment}

We do not consider in this paper the differences induced by solenoidal versus compressible driving of turbulence (see \citealt{2021A&A...656A.118S}). For incompressible fluids that we deal with, the latter driving is impossible. We note that our considerations that justify Eq. (\ref{3}) can also apply to the compressible case if the flow induces $\delta B_\|$. Note that numerical simulations in \citealt{2021A&A...656A.118S} support Eq. (\ref{3}) for compressible MHD simulations. However, a more detailed discussion of the effects present in compressible sub-Alfvenic turbulence is beyond the scope of this paper. 

By dividing both parts of Eq. (\ref{2}) by $B$ and accounting for the definition of Alfven velocity $V_A=B/\sqrt{4 \pi \rho}$, one gets
\begin{equation}
    \delta B \approx \sqrt{\rho} V_L \left(\frac{V_L}{V_A} \right),
    \label{4}
\end{equation}
where we used $\delta B=\sqrt{B_\|^2+B_\bot^2}$ accounted for Eq.(\ref{3}). Eq. (\ref{4}) shows that for velocity driving, the fluctuations of magnetic field and velocity at the injection scale are not related through the Alfven relations, but kinetic energy dominates for subAlfvenic velocity driving. This has significant implications that we discuss further in the paper.

A peculiar feature of our discussion above is that we did not separate the magnetic perturbations in fundamental MHD modes, i.e., slow, Alfven, and fast. Instead, we deal with parallel and perpendicular magnetic field fluctuations $\delta B_\|$ and $\delta B_\bot$, which, as we discuss in Appendix A, can have a different way of decomposition into fundamental modes in media with different sound $v_s$ and Alfven velocities $V_A$  ratios. Therefore, our estimates may apply to compressible simulations as discussed in \S \ref{sec:beattie20}. The sound velocity is much larger than the Alfven velocity in the weakly compressible fluid, $v_s\gg V_A$. In this case, the fluctuations of velocity moving along the magnetic flux tube induce variations in the diameter that can be associated with slow modes. The same Bernoulli effect in the flux tube for $V_A\gg v_c$ induces the variations of flux tubes described by fast modes. In Appendix A we explain this in more detail how $\delta B_\|$ is produced by slow and fast modes in a more general setting. Both effects are described by Eq. (\ref{2}), i.e., the magnetic response of the media to sub-Alfvenic velocity driving does not depend on fluid compressibility. However, the case of incompressible MHD turbulence is simpler and more fundamental. Thus, for the rest of the paper, we will focus on it. 

In the incompressible case, the extent of the magnetic field perpendicular dilatations, i.e., $L M_{A, b}$, determines the perpendicular coherence scale of the slow mode wavepackets. Further in the paper, we discuss the consequences of this effect.

\subsubsection{Relation to strong turbulence}

For our discussion, it is important to keep in mind that non-linearity is a property of fluids, not magnetic fields.
The lessons that one can extract from our discussion of strong turbulence above are as follows: 
\begin{enumerate}
\item In MHD, turbulence at small scales evolves through non-linearities of hydrodynamic equation term $(v\nabla)v\sim v_\bot^2/l_\bot$, where it is taken into account that magnetic field constrains the parallel fluid motions. This results for in the non-linear rate of evolution $\omega_{nl}\sim v_\bot/l_\bot$.  The non-linearity comes from hydrodynamics with magnetic fields acting to restrict the possible fluid motions; 
\item The non-linear hydrodynamic term decorelates motions within eddies, inducing the cascade of energy towards smaller scales\footnote{This can be seen as a process of higher Fourier harmonics generation by the non-linear term.}; 
\item The critical balance given by $\omega_{nl}\sim t_A^{-1}$ (see Eq. (\ref{balance}) ) is the essential part of a strong turbulence regime. This induces the limit for the perpendicular scale of motions for strong turbulence given by Eq. (\ref{ttr}). 
\end{enumerate}
Extending this reasoning to the scales  $[l_{tr,\bot}, L]$, we have to abandon point (3) above,  as the cascading rate $\omega_{nl}$ is smaller than propagation rate $t_A^{-1}$. This is the point for which the physical picture that we considered earlier must be modified.

We may approach the problem of {\it weak Alfvenic turbulence} more formally, considering the 3-wave interaction of waves with frequencies $\omega$ and wavenumber ${ k}$. The conservation of energy defined as $\omega_1 +\omega_2=\omega_3$ should be satisfied together with the conservation of the momentum defined as ${ k_1}+ { k_2}= { k_3}$  (see \cite{1996ApJ...465..845N}). Accounting for the dispersion relation $\omega=\pm k_\| V_A$, one has to accept that in the weak regime, there is no change of $k_\|$, but only the change of $k_\bot$ can occur. This means that the frequency of the waves in weak turbulence $\omega \sim k_\| V_A\sim V_A/l_\|$ does not change and stays equal to the frequency of waves at the injection scale $L$. 
This means that the cascading, i.e., the decorrelation of motions and higher spatial harmonics, appears only perpendicular to the magnetic field. This cascading happens over several $t_A$. \footnote{We note that the representation of wavenumbers, i.e., Fourier components, is adequate for weak turbulence for which perturbations are defined in the mean-field reference frame (see \citealt{CV00} for a discussion). Thus, one can use both $l$ and $k$ notations interchangeably for weak turbulence.}

Like strong turbulence, weak turbulence is induced by motions perpendicular to the magnetic field. The weak turbulence cascade creates the pattern of filaments rotating perpendicular to the magnetic field. As these filaments get thinner than $l_{tr,\bot}$, the perpendicular fluid motions induce Alfvenic perturbations with frequency $\sim v_l/l_\bot$, invalidating the weak turbulence requirement of constant $\omega$. This is the regime of strong turbulence that we described with the eddies. In other words, strong turbulence is a natural continuation of the weak cascade for the scales at which the magnetic field tubes get insufficiently rigid.

\subsubsection{Velocity scaling}

\noindent{\it Perpendicular velocity scaling. ---}
The non-linear hydrodynamic term $\sim v_\bot/l_\bot$ induces cascading for the fluid moving along perturbed magnetic field lines. As we discussed earlier, the turbulent motions of weak Alfvenic turbulence are exclusively the motions perpendicular to the magnetic field while that $\omega$ does not change. The non-linear hydrodynamic term distorts the original velocity pattern and creates more structure perpendicular to the magnetic field direction. The acceleration caused by this term is $v_\bot\omega_{nl}$ acts over the period of motion driving $\omega^{-1}$.  This induces a change of the momentum of the fluid element for incompressible fluid is  proportional to $v_\bot \omega_{nl}/\omega\sim v_k\chi$, where 
\begin{equation}
    \chi=\frac{(v_\bot/l_\bot)}{\omega}.
    \label{chi}
\end{equation}
The restrictive role of the magnetic field is clear from Eq. (\ref{chi}). The stronger the magnetic field, the larger $\omega$, the smaller the non-linear momentum change.

For weak turbulence, $\chi\ll 1$, meaning that magnetic field strongly inhibits non-linearities. Thus, many non-linear interactions are required to significantly decorate/cascade the velocity motions to smaller scales. In turbulence, the interactions are incoherent, which induces the random walk of the non-linear distortion
\begin{equation}
    \delta v^2 \approx v_\bot^2\chi^2 N,
    \label{random}
\end{equation}
where $N\approx t\omega$ is the number of steps performed in time $t$. When the change of squared momentum $\delta v^2$ gets of the order of the original $\sim v_\bot^2$, the motions are randomized. This requires $N\sim \chi^{-2}$ steps.

This condition can be used to define the cascading time $\tau_{cas}$ of weak turbulence, i.e.
\begin{equation}
    \tau_{cas,weak}^{-1}\approx \frac{\omega}{N} \approx\frac{(v_l/l_\bot)^2}{\omega} ,
    \label{cascade}
\end{equation}
where for the uniformity of the presentation with the case of strong turbulence, we used $l_\bot \sim k_\bot^{-1}$. Note that the cascading time given by Eq. (\ref{cascade}) is well-accepted in the MHD turbulence literature (see \citealt{Kra65}). We note that even though Eq. (\ref{random}) does not apply formally to strong turbulence, Eq. (\ref{cascade}) smoothly transfers to provide the correct cascading rate.\footnote{For strong turbulence $\omega\sim V_A/l_\|$ changes with the scale $l_\|$ and provides $\tau_{cas, strong}\sim v_l/l_\bot$ due to Eq. (\ref{cascade}). This coincides with the hydrodynamic rate of eddy turnover around the direction of the magnetic field. The possibility of describing the strong turbulence as both eddies and waves corresponds to the duality of MHD turbulence nature discussed in \cite{CL03}. }

The velocity scaling in weak turbulence can be derived trivially by noting that for weak turbulence $\tau^{-1}_{cas, weak}\ll \omega$ due to Eq. (\ref{cascade}). As $l_\|$ does not change in the weak regime, $\omega$ stays constant. Therefore, the cascade of kinetic energy for $l_\bot$ in the range $[l_{tr,\bot}, L]$ can be described for incompressible motions as
\begin{equation}
    v_{l,weak}^2/\tau_{cas, weak}\sim \frac{v_l^4}{l_\bot^2\omega}\sim const
    \label{energycascade}
\end{equation}
which provides for weak MHD turbulence, i.e., for constant $\omega$, the velocity scaling (LV99):
\begin{equation}
    v_{l, weak}\approx V_L \left(\frac{l_\perp}{L}\right)^{1/2}.
    \label{weak_vel}
\end{equation}
This scaling of velocity induces the scaling of kinetic energy $E_k \cdot k_\perp \sim v_k^2$, which results in 
\begin{equation}
    E_k \sim k^{-2}_\bot,
    \label{Ekbot}
\end{equation}
which is an accepted expression for the kinetic energy spectrum for weak turbulence (see LV99, \citealt{2000JPlPh..63..447G}). 

Consider a toy model with a turbulence driving force $\delta F$ being random in space. The increase of kinetic energy in a parallel direction is $\delta F \delta l$, with $\delta$ being the transposition of fluid over the time $\delta t$, i.e., $\delta l \sim a (\delta t)^2$, where the acceleration in parallel to magnetic field direction $a$ is $\sim \delta F/\rho$. The injection velocity is $V_l\sim a \delta t$ in that direction. The transposition in the perpendicular to magnetic field direction is $M_{A,b}$ times smaller due to the backreaction of the magnetic tension. As a result, the energy induced by the isotropic driving in a perpendicular direction is reduced by a factor $M_{A,b}$ compared to the parallel direction. In other words, the amplitude of the kinetic energy associated with parallel to magnetic field motions ${\cal E_\|}$ is expected to be $\sim M_{A,b}^{-1}$ times the kinetic energy associated with perpendicular to magnetic field motions ${\cal E_\bot}$. 

This agrees with the conclusion about the sub-dominance of the energy of magnetic turbulence fluctuations that follows from Eq. (\ref{4}). The cascade starts at the injection scale $L$ for the isotropic driving.

\noindent{\it Fluctuations in the flow parallel to the magnetic field. ---} 
Fluid motions along magnetic field lines provide minimal resistance for isotropic velocity driving. Thus, as we discussed earlier, most energy is transferred this way.  If the magnetic field were laminar, such motions could induce only acoustic-type turbulence. We disregard this possibility as acoustic turbulence does not exist in the incompressible limit we consider. 

In the incompressible case, the structures created by the parallel fluid flow correspond to slow mode perturbations with the parallel magnetic field size $L$ and perpendicular scale $L M_{A,b}$. These structures are sheared by Alfvenic fluctuations discussed above. This transfers the scaling of weak Alfven modes given by Eq.(\ref{flow}) to slow modes but does not cascade the energy in the parallel direction.

In the case of fluid moving along a turbulent magnetic field that has the structure in the parallel direction, this structure gets imprinted to the flow. In other words, the magnetic field stochasticity is transferred to the stochasticity of the conducting fluid flow. As discussed in \S \ref{scal}, strong Alfvenic turbulence introduces the structure in the parallel direction. Thus, following magnetic field lines, the velocity fluctuations reflect scaling of Alfvenic fluctuations parallel to the magnetic field, which (see \S \ref{scal} have the energy spectrum $k_\|^{-2}$. Therefore, the fluid flow along magnetic field lines exhibits parallel fluctuations 
\begin{equation}
    E_{k}\sim k^{-2},
    \label{flow}
\end{equation}
We expect this spectrum to change for $k>1/LM_{A,b}$, as the strong MHD turbulence scaling dominates the spectrum at small scales. In particular, the fluid motions along magnetic field lines become a part of the pseudo-Alfven cascade. 

This stochastic fluid motion is not a part of MHD turbulence driven through magnetic fluctuations. It represents an additional element introduced by the velocity driving. 

\subsubsection{Scaling of Alfvenic fluctuations }

Magnetic field scaling in weak turbulence can present several regimes depending on the turbulence driving. If the driving is "narrow frequency driving," i.e., $\delta \omega\ll \omega_L$, the driving frequency $\omega_L$ will determine the principal frequency of parallel fluctuations in the numerical box. If the corresponding wavelength of the induced fluctuations $V_A/\omega_L$ is larger than the box size $L$, the large-scale magnetic fluctuations cannot be observed on the box size scale, naturally resulting in the dominance of the turbulent kinetic energy compared to the magnetic energy.

The situation is different if
the driving is "broadband."
The critical scale for which the driving excites the Alfvenic perturbation that fits the numerical box corresponds to the wave frequency:
\begin{equation}
    \omega_{c}=\omega_L \delta B/B,
\end{equation}
For frequencies larger than $\omega_c$, the turbulent cascade is Alfvenic, with kinetic and magnetic energies equal at such scales. 

\noindent{\it Injection coherence scale. ---}  As we mentioned in \S \ref{2.2.1}, for isotropic velocity driving, one can introduce the coherence scale corresponding to the perpendicular scale of the wavepackets. The velocity fluctuations injected at scale $L$ propagate $LM_{A,b}$ perpendicular to the magnetic field. Thus,
the perpendicular scale of fluctuations is limited by
\begin{equation}
    L_{\bot, max}\approx LM_{A,b},
\end{equation}
%{\color{red} We must add Ethan's statement in %abstract back here.}
which correspond to the perpendicular fluctuations arising from the largest scale driving. This corresponds to the upper boundary for the motions correlated perpendicularly. The cascade gets into a strong turbulence regime at the scale $l_{tr}=LM_{A,b}^2$. The velocities and magnetic field are related through the Alfven relation. According to Eq. (\ref{vl}), the velocity amplitude at this scale is $V_L M_A$. 
The increase of velocity and magnetic field fluctuations with the scale in the range $[l_{tr}, L_{\bot, max}]$ follows the weak MHD cascade given by Eq.(\ref{weak_vel}). Thus, the maximal amplitude of Alfvenic perturbations corresponds to 
\begin{equation}
    \delta B_{max}\approx V_L\sqrt{4\pi \rho} M_{A,b}^{1/2},
\end{equation}
which suggests that the Alfvenic turbulence energy is $\sim M_{A,b}$ fraction of the kinetic turbulent energy. We will see further that the amplitude $\delta B_{max}$ corresponds to our earlier estimate of the magnetic fluctuation from Eq. (\ref{4}) if one accounts for Eq. (\ref{M^2}).

\subsubsection{Scaling of parallel to magnetic field fluctuations}

Parallel to magnetic field fluctuations for incompressible flow are pseudo-Alfven fluctuations, which are the limiting case of slow modes.  The importance of slow modes for explaining unusual properties of sub-Alfvenic simulations was stressed by Chris McKee (private communication). These compressible perturbations propagate with Alfven speed $V_A$ ({see also Appendix A). 

Eq. (\ref{2}) describes the generation of pseudo-Alfven modes by fluid flows along magnetic field lines.  
The shearing of pseudo-Alfven fluctuations by the Alfvenic cascade is efficient when the Alfven shearing rate $v_l/l_\bot$ becomes equal to the propagation rate of pseudo-Alfven fluctuations over the injection scale, i.e., $V_A/L$. Equating the two, one gets the transition scale $l_{tr}$ given by Eq. (\ref{ttr}). In other words, the cascading of pseudo-Alfven fluctuations by Alfvenic cascade starts from perpendicular scale $LM_{A,b}^2$, which is smaller than the scale for the initiation of the Alfven cascade through velocity driving by a factor $M_{A,b}$.

As we discussed earlier, in the case of velocity driving, the amplitude of pseudo-Alfven fluctuations according to Eq. (\ref{2}) is $V_L M_{A,b}^{1/2}$. The direct non-linear interaction of such fluctuations results in their cascading being slow. Thus, the pseudo-Alfven fluctuations $\delta B_s$ stay constant over the scales $[l_{tr}, L]$. The condition $\delta B_s\approx const$ results in the spectrum $\delta B_s \delta B_s\sim E_{B,s}k$ with scaling of the pseudo-Alfven modes\footnote{ In our next paper (Yuen et. al in prep) we consider in detail the properties of the parallel and perpendicular spectra for these fluctuations for different types of driving.}:
%{\color{red} KH:it is actually $k_\perp^0$ in large scale. Below is true only in small scale (1D fluc is always decoherent). Moreover, has to emphasize below is correct only in isotropic driving.}
%{\color{blue} AL: We are dealing only with isotropic driving. Other spectra we can refer to your paper.}
% KH: To be exact, we were considering also isotropic driving as in our paper. 
\begin{equation}
    E_{B, s}(k) \sim k^{-1}.
    \label{EBs}
\end{equation}

With velocity evolving according to Eq. (\ref{Ekbot}) and the magnetic energy evolving according to Eq. (\ref{EBs}), the difference between the two energies at the injection scale $L$ amounts to   
\begin{equation}
    E_{B,s} (k_L) \approx E_k (k_L) M_{A,b},
    \label{K/B}
\end{equation}
where $k_L\sim 1/L$, which demonstrates the self-consistency of our picture. 

It follows from Eq. (\ref{2}) that the isotropic velocity driving is expected to induce the $\delta B_\bot\sim \delta B_\|$  at the scale of injection. Thus, we expect that:
\begin{equation}
    E_{B,s} (k_L)\approx E_{B,A} (k_L)
\end{equation}

The Alfvenic cascading is expected to operate at the perpendicular scale $LM_{A,b}$. At the $LM_{A,b}^2$ scale, the amplitude of Alfvenic fluctuations is reduced by a factor $M_{A,b}^{1/2}$, which suggests the dominance of energy of pseudo-Alfven/slow modes over Alfven modes. A rough estimate of the relative amplitudes is
%{\color{red} KH: transition happens at $LM_{A,b}$ and the below relation applies only to $k_\parallel=0$, see condensate Fig 1. Also, A/S is in equi-partition as long as 0 mode is removed and $k_\perp$ big. The below equation is essentially taking averages over the number of legs between hens and rabbits, and claims that the number of legs for the creatures on this planet is 3.}
%{\color{blue} Let us not deal with rats and rabbits in this paper. Condensate is a separate publication. I looked through it and have my issues with it.}
\begin{equation}
    E_{B, A} (k)\approx E_{B, s} (k) M_{A,b}^{1/2}, ~~~~k>\frac{1}{l_{tr}},
    \label{slow_alf}
\end{equation}
where $E_{B,A}$ is the spectrum of Alfven modes, $l_{tr}$ is given by Eq. (\ref{ttr}).
 Additional effects for $M_{A,b}<1$ turbulence are related to the effect of accumulation of energy at the $k_\|=0$ modes, which is frequently referred to by plasma turbulence community as "condensation" process (see \citealt{2012PhRvE..85c6406S}). We consider these effects in the subsequent paper (Yuen et al. 2024, in prep).

How stiff the magnetic field lines are in relation to the flow depends on the forcing scale and $M_A$. The magnetic field lines stay stiff for the motions $l_\| >l_\bot M_{A,b}^{-1}$. Such motions induced by the driving induce velocity fluctuations rather than magnetic fluctuations.

\subsection{Two forms of Alfven Mach number}
\label{sec:2.3}

We note that the accepted textbook definition of the "velocity Alfven Mach number" is the ratio of the turbulence injection velocity to the Alfven one, i.e.
\begin{equation}
    M_A\equiv \frac{V_L}{V_A}.
    \label{Alfven}
\end{equation}

Our study reveals that for the velocity-driven {\it incompressible} sub-Alfvenic turbulence, the kinetic energy exceeds the magnetic turbulent energy.  Eq. (\ref{K/B}) corresponds to $\delta B^2 \sim \rho V_L^2 M_A^2$. By dividing both parts by $V_A^2$, we can see that the "magnetic Alfven number" $M_{A,b}$ is related to the "velocity Alfven number" $M_A$ as
\begin{equation}
   M_{A,b}=\frac{\delta B}{B}\approx M_A^2,
    \label{M^2}
\end{equation}
which is in contrast to the textbook definition given by Eq. (\ref{MA,b}).
The same result follows from dividing both parts of Eq. (\ref{4}) by $B$.
The properties of the most important part of MHD turbulence, the Alfvenic cascade, depend on $M_{A,b}$, which is a fundamental number. 

Note that the theory of MHD turbulence is traditionally formulated for incompressible fluid settings where only Alfven and pseudo-Alfven fluctuations are considered (see GS95). It is assumed that velocity fluctuations and magnetic field fluctuations are related through the Alfven relation. However, as we showed, this is not the case for velocity-driven subAlfenic turbulence (see Eq. (\ref{4}). 

 Our paper deals with the properties of incompressible MHD turbulence. The fluid compressibility introduces an additional complexity. All three modes, Alfven, slow, and fast, contribute to $V_L$ for compressible MHD turbulence. The properties of turbulence depend on the relative contribution of the modes, apart from the value of $V_L$. This effect explains the difference in turbulence properties for solenoidal and potential driving obtained for the same $M_A$ (see \citealt{2008ApJ...688L..79FZ}).

We note that the 
relation between $\delta B/B$ and $M_A$ given by Eq. (\ref{M^2}) was reported in simulations \citep{2020MNRAS.498.1593B} and attributed to the effects of fluid compressibility. Similarly, by combining Eq. (\ref{4}) and Eq. (\ref{Alfven}), one can get the relation between kinetic and magnetic turbulent energies $E_B\approx E_k M_A^2$, which corresponds to the result established through the analysis of compressible numerical simulations \citep{2022MNRAS.515.5267B}.\footnote{Another way to obtain this relation is to combine Eq. (\ref{K/B}) with Eq. (\ref{M^2}).} Our study suggests that, in fact, these effects arise from velocity driving of MHD turbulence. We plan to numerically test this claim elsewhere but provide more physical arguments supporting our point in \S \ref{sec:beattie20} and in Appendix A. 

\section{Numerical simulations}
\label{sec:num}
In this study, we perform the incompressible MHD simulations with the pseudo-spectral code called MHDFlows.jl \footnote{https://github.com/MHDFlows/MHDFlows.jl} \citep{MHDFlows}. MHDFlows is the newly developed MHD open source code based on the dynamical language Julia with spectral solver framework FourierFlows.jl \citep{FourierFlows}. In contrast to the traditional spectral solver, it supports native GPU acceleration and supports wide ranges of MHD simulations, including ideal MHD, Electron-MHD, and MHD with volume penalization method. In this study, we solve the ideal MHD equation in the periodic box with the size of $2\pi$: 

\begin{equation}
\begin{aligned}
\frac{\partial \vec{v} }{\partial t} + (\vec{v} \cdot \nabla )\vec{v} &= -\nabla P + (\nabla \times \vec{B}) \times \vec{B} + \nu \nabla^2 \vec{v} 
 \\
\frac{\partial \vec{B} }{\partial t} &= \nabla\times (\vec{v}\times\vec{B}) + \mu \nabla^2 \vec{B}
\end{aligned}
\label{eq:eom}
\end{equation}
{All the symbols have their usual meaning. We use the method mentioned in \cite{2001ApJ...554.1175M}, in which pressure P is chosen such that the equations maintain the divergence-free condition throughout the simulation. The incompressibility of spectral code is a well-known property.
Figure \ref{figure1_test} demonstrates that the divergence level and the compressible term is magnitude are observed to be on the order of $10^{-6}$, which is of the order relative error limit of Float32 format used in the computation. Uncertainties of such order cannot compromise the incompressible nature of our code to induce any noticeable effects on the time of our simulations.} The turbulence is driven on a large scale (k=1.5) through the method proposed by \cite{A99}. In addition, a seed field with different strengths was injected at the beginning of the simulation to control the $M_A$, and we analyzed the result after ten large-scale eddy turnover times to wait for the saturation of kinetic and magnetic energies of turbulence. We use the Runge–Kutta method of order 4 (RK4) for the time integration and 2/3 as the aliasing factor (See \citealt{1971JAtS...28.1074O}). Table \ref{tab:sim} shows the key parameters of our simulations. To characterize the magnetization of our simulations we use the "velocity-based" definition of the Alfven Mach number $M_A$ given by Eq. (\ref{Alfven}). 

\begin{figure}
\centering
\includegraphics[width=1\linewidth, trim=0 135 0 135, clip]{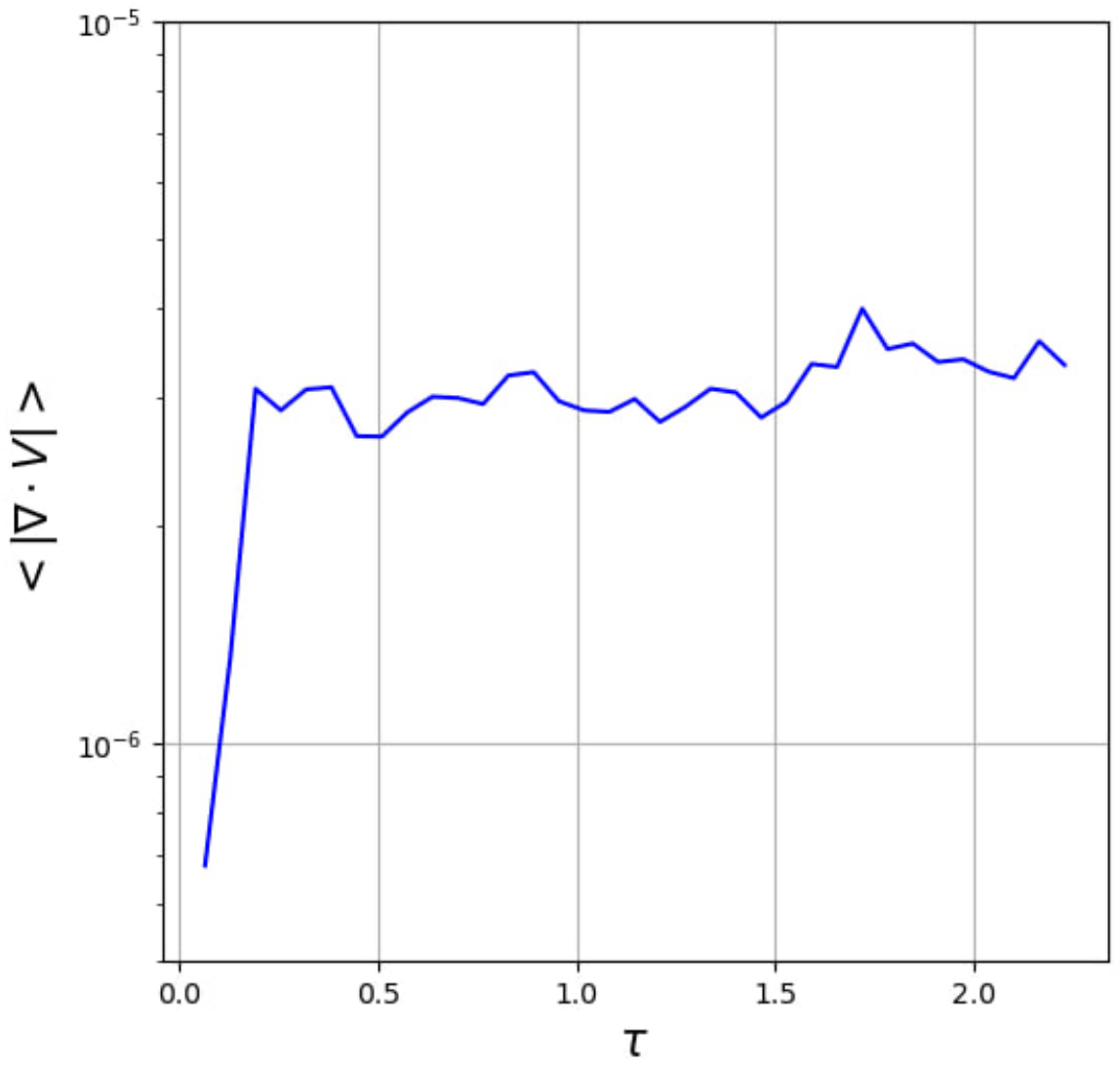}
    \caption{The absolute value of the compressible term in our code (Simulation used: M1).}
    \label{figure1_test}
\end{figure}

\begin{table}
\centering
\begin{tabular}{@{}lllll@{}}
\toprule
Simulation & \( \langle M_A \rangle \) & \( N^3 \)              & \( \eta \& \mu  \)                   \\ \midrule
M0         & 0.393                     & \multirow{10}{*}{\(384^3\)} & \multirow{10}{*}{\(5 \times 10^{-5}\)} \\ \cmidrule(r){1-2} 
M1         & 0.451                     &                           &                                   \\ \cmidrule(r){1-2} 
M2         & 0.588                     &                           &                                   \\ \cmidrule(r){1-2} 
M3         & 0.685                     &                           &                                   \\ \cmidrule(r){1-2} 
M4         & 0.825                     &                           &                                   \\ \cmidrule(r){1-2} 
M5         & 0.904                     &                           &                                   \\ \cmidrule(r){1-2} 
M6         & 0.906                     &                           &                                   \\ \cmidrule(r){1-2} 
M7         & 0.918                     &                           &                                   \\ \cmidrule(r){1-2} 
M8         & 1.041                     &                           &                                   \\ \cmidrule(r){1-2} 
M9         & 1.264                     &                           &                                   \\ \bottomrule
\end{tabular}
\caption{key parameter of the MHD incompressible simulation will be used in this paper.$\langle M_A\rangle,\eta=\mu,N^3$ refer to the 10-slice time-averaged Alfvénic Mach Number, kinetic/magnetic diffusivity, and resolution of the simulation, respectively. The magnetic Prandtl number is 1 for all simulations. The time-averaged Alfvénic Mach Number is computed through the last 5 snapshots of each simulation.}
\label{tab:sim}
\end{table}

\section{Analysis of numerical results}
\label{sec:analysis}
\subsection{Kinetic and magnetic energy of turbulence}

Figure \ref{visual1} provides the visualization of our simulations corresponding to $M_A\approx 0.45$. 
The initial direction of the magnetic field is along the $z$-axis.
The magnetic field and velocity are shown, and in our  incompressible MHD simulations with velocity driving, we observe that the amplitude of velocity fluctuations exceeds those of the magnetic field.\footnote{We note, parenthetically, that a similar difference was observed in earlier numerical studies of {\it compressible} MHD turbulence \citep{2022MNRAS.515.5267B, 2023A&A...672L...3S} that also employed velocity driving.  This suggests that the disparity in magnetic and velocity turbulent energies {\it is not} arises from velocity driving rather than effects related to fluid compressibility.} We also observe that the magnetic field is more structured than velocity. This indicates a more shallow energy spectrum of the magnetic field at the scales close to the injection scale.   Statistically, this alignment is preserved in the driving process, but different realizations that we analyze exhibit the deviations of the mean field from its initial direction.

\begin{figure*}
\centering
\includegraphics[width=1\linewidth]{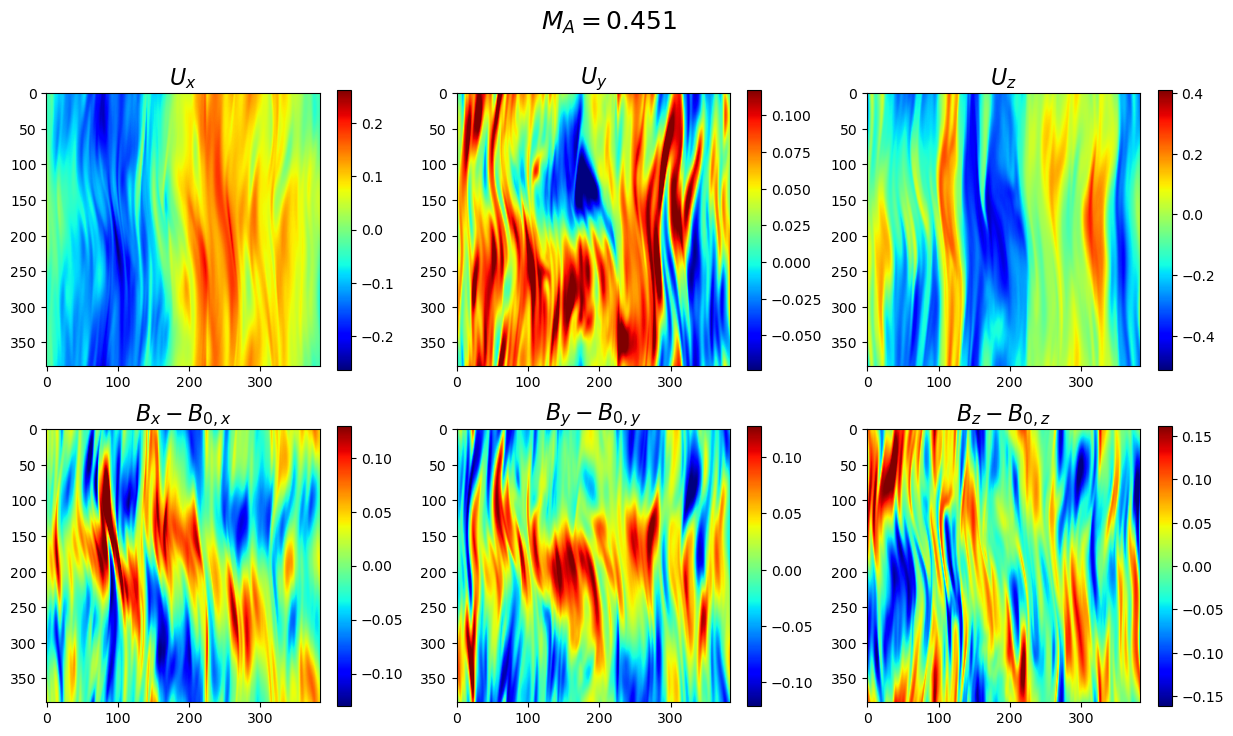}
    \caption{2D slice of simulations with $M_A\approx 0.45$. The mean field directed along the z-axis was subtracted to focus on fluctuations. Upper raw are velocity components, the lower raw are the components of magnetic fluctuations.}
    \label{visual1}
\end{figure*}

Figure \ref{spectrum1} shows spectra of the magnetic field and velocities for different $M_A$. The magnetic field spectrum for large scales is consistent with the predicted magnetic energy spectrum $E_B\sim k^{-1}$ (see Eq. (\ref{EBs}), while the kinetic energy spectrum corresponds to, i.e., $E_{k}\sim k^{-2}$, which agrees with our prediction given by Eq.(\ref{flow}).

\begin{figure*}
\centering
\includegraphics[width=1\linewidth]{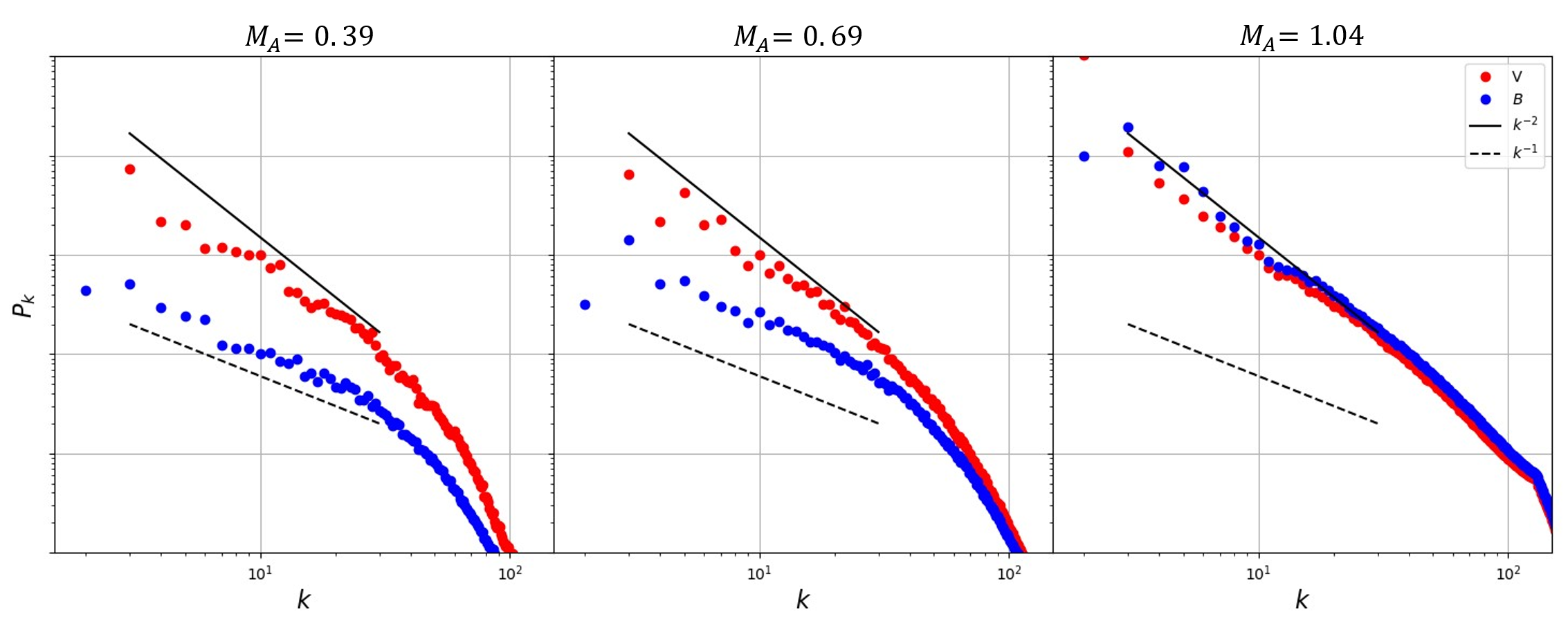}
    \caption{Velocity and magnetic field spectra for different $\delta B/B$, showing the change of magnetic spectra. The disparity between magnetic and velocity turbulent energies is large for small $M_A$. When magnetic fluctuations become of the order of unity, the disparity between kinetic and magnetic energy disappears. The solid and dashed lines correspond to the theoretical expectations for velocity and magnetic spectra for subAlfvenic turbulence given by Eq.(\ref{flow}) and Eq. (\ref{EBs}). }
    \label{spectrum1}
\end{figure*}

The convergence of the spectra around the value $l_{tr}$ is observed. As the $M_A$ increases, the difference between the magnetic and velocity spectra decreases, as shown in Figure \ref{spectrum1}, with two spectra merging towards the GS95 incompressible turbulence spectrum corresponding to $M_A=1$. On the contrary, as $M_A$ decreases, the difference between the velocity and magnetic field spectra gets more prominent. Figure \ref{spectrum1} shows the spectra for $M_A\approx 0.09$. For such low $M_A$, turbulence reaches the dissipation scale before its transfer to the strong MHD turbulence regime. 

To test whether the disparity of magnetic and kinetic energies arises from the way we drive turbulence, in Figure \ref{fig:B-field_Driving}, we provide spectra of sub-Alfvenic turbulence driven by magnetic fluctuations. In this case, we observe that $E_k=E_B$ corresponds to our theoretical expectations.

\begin{figure}
\centering
\includegraphics[width=1\linewidth]{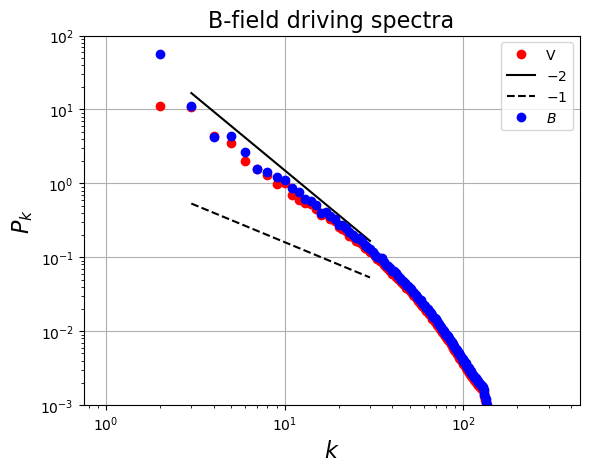}
    \caption{Energy spectra for both velocity and magnetic field with magnetic field driving of sub-Alfvenic turbulence, with $M_A=0.7$ In contrast with the velocity driving, the magnetic and kinetic energies are at equi-partition.}
    \label{fig:B-field_Driving}
\end{figure}

To test to the effect of the resolution, we also perform a convergence test of the magnetic spectrum with different resolutions in Figure \ref{fig:resolution}, showing spectra converged for all resolutions. We realize that to claim to establish the exact spectral slope numerically, one requires a separate dedicated study. Thus, we cautiously use "consistent" to describe our numerical results.

%Our major prediction is the change of the kinetic to magnetic energy ratio given by Eq. (\ref{K/B}).The dependence of this ratio on the value of the magnetic field fluctuation is shown in Figure \ref{ratio}. The observed linear dependence corresponds to Eq. (\ref{deltaphi}).

\begin{comment}
\begin{figure}
\centering
\includegraphics[width=1\linewidth]{ratio.png}
    \caption{The ratio of time-averaged magnetic to kinetic energies as a function of the amplitude of the time-averaged normalized magnetic fluctuation $\delta B/B$.}
    \label{ratio}
\end{figure}

The measurements of magnetic fluctuations parallel and perpendicular to the mean magnetic field presented in Figure \ref{ratio2} show that their ratio does not change much with the Alfven Mach number. This corresponds to the our theoretical considerations suggesting that the magnetic fluctuations induced by Alfven and slow modes should be approximately equal. 

\begin{figure}
\centering
\includegraphics[width=1\linewidth]{ratio2.png}
    \caption{The ratio of perpendicular to parallel fluctuations as a function of the amplitude of the normalized magnetic fluctuation $\delta B/B$.}
    \label{ratio2}
\end{figure}
\end{comment}

\begin{figure}
\centering
\includegraphics[width=1\linewidth]{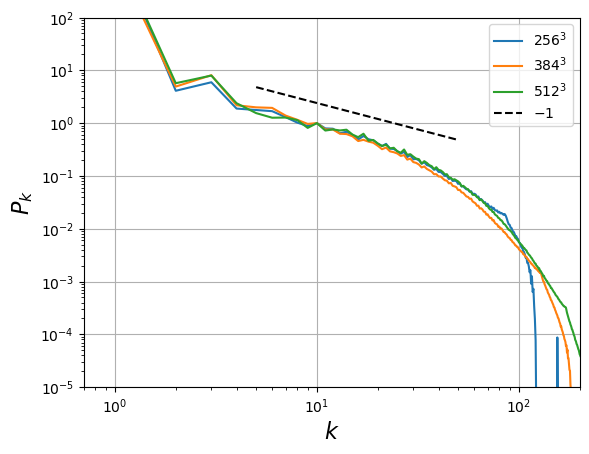}
    \caption{Comparison of magnetic spectra obtained with different resolutions using the setting of M3 corresponding to $M_A\approx 0.28$. The spectrum $E_B\sim k^{-1}$ corresponding to $\delta B=const$ is shown by a dashed line. }
    \label{fig:resolution}
\end{figure}

%A non-trivial result of our study is that in incompressible media for $M_A<1$ the dispersion of magnetic field $\delta B$ is proportional to $M_A^2$. This is a direct consequence of the stagnation of the fluctuation of $\delta B$ at the value obtained at the end of the strong turbulence cascade. We test this prediction in Fig. (\ref{stagnation})

\subsection{Parallel and perpendicular fluctuations}

The kinetic and magnetic energy spectra measurements do not capture the difference between the motions along and perpendicular to the magnetic field. Such measurements must be done with respect to the magnetic field's local direction. Therefore, the spectra measured in the frame of the mean magnetic field are not sufficiently informative. To deal with the problem,  we employ the second-order structure functions measured with respect to the local magnetic field for our study (see \citealt{CV00}). For such measurements, the magnetic field of the magnetized eddy defines the reference frame.

To stress that the measurements are done in the local system of reference in Figure \ref{fig:sf_temporal}, we use the word "local" in front of structure functions of magnetic and velocity field. Figure \ref{fig:sf_temporal} demonstrates that both the fluctuations of the magnetic field and the velocities are elongated along the magnetic field direction, but their anisotropy is very different. This is a clear distinction from the strong MHD turbulence, where the velocities and magnetic field exhibit the same scaling, as discussed earlier. In the sub-Alfvenic velocity-driven case that we consider, the anisotropy of velocities is more prominent compared to the magnetic fluctuations. We expect this effect to be slightly mitigated in compressible simulations due to the presence of fast modes. Note that the strong anisotropy of velocity fluctuations is observed in compressible  turbulence simulations in \cite{2022ApJ...928..112B} for $M\ll 1$. The velocity structure functions are strongly correlated with the direction of the magnetic field due to the fluid flowing along the magnetic field. 

We show both structure functions corresponding parallel fluctuations, i.e., $SF_{2B}(k_\|=0)$ and $SF_{2V}(k_\|=0)$, as well as perpendicular fluctuations, i.e., $SF_{2B}(k_\bot=0)$ and $SF_{2V}(k_\bot=0)$. We only use $k_\bot$ and $k_\|$ as shortcut convenient notations. The measurement is done in the local system of reference, which differs, in general, from the mean-field magnetic field reference frame.
The results are strikingly different for small $M_A$, indicating the different physics of motions parallel and perpendicular to the magnetic field.

The energy of magnetic compressions corresponding to pseudo-Alfven modes (see \citealt{CL03}) is subdominant at large scales compared to kinetic energy.

\begin{figure*}
\centering
\includegraphics[width=0.98\linewidth]{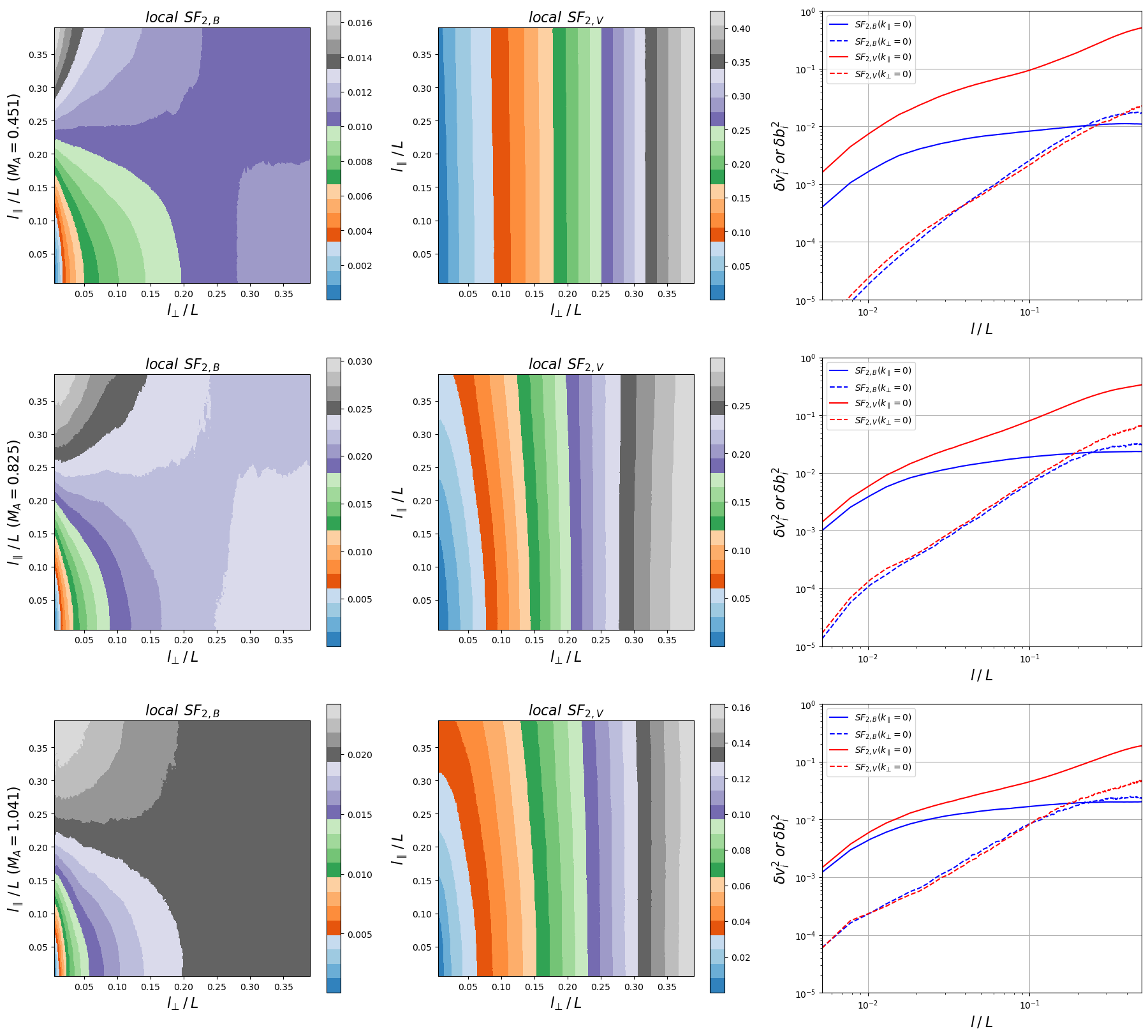}
    \caption{ Left: Contours of Second Order Local Structure function of magnetic field $SF_{2,B}$, Mid: Contours of Second Order Local Structure function of velocity field $SF_{2,V}$. Right: Variation of parallel and perpendicular structure functions for different $M_{A}$. The dominance of parallel velocity fluctuations increases with the decrease of $M_A$.}
    \label{fig:sf_temporal}
\end{figure*}

%The properties of perpendicular to magnetic field fluctuations are different. The amplitude of velocity fluctuations perpendicular to the magnetic field is smaller than the amplitude of the fluctuation parallel to the magnetic field with the energy relation $E_{\k,\bot}\approx E_{k,\|}M_A^{?}$. { please check}. However, this difference is unimportant for the general energy balance dominated by the parallel velocity fluctuations. 

For small $ M_A$, the amplitude of velocity fluctuations parallel to the magnetic field significantly exceeds the one perpendicular to the magnetic field.  The latter is similar to the amplitude of magnetic fluctuations in the units of Alfven velocity, indicating that the fluctuations perpendicular to the magnetic field are connected by the Alfven relation. In contrast, the amplitude of velocity fluctuations is larger than that of the magnetic fluctuations in Alfven velocity units, demonstrating that velocity fluctuations along the magnetic field dominate. We associate those with flows along magnetic field lines induced by velocity driving. This effect is absent for the magnetic driving of subAlfvenic turbulence.

Another significant effect illustrated by Fig. \ref{fig:sf_temporal} is the higher amplitude of magnetic compressible fluctuations,  i.e., pseudo-Alfvenic fluctuations,  compared to Alfven fluctuations at small scales. The amplitude of the fluctuations is similar at the injection scale. Still, the compressible fluctuations marginally change their amplitude at large scales, while the Alfvenic perturbations follow the turbulent cascade for all scales. This corresponds to our expectations in \S\ref{sec:weakturb}. As we discuss later, this newly identified effect has astrophysical significance.

\subsection{Simulations of synthetic observations for two ways of driving}

There is a significant difference between the results obtained with magnetic and velocity driving of turbulence. We provide the synthetic observations of magnetically and velocity-driven (Fig. \ref{fig:both_figures}) MHD cubes to test how to distinguish the two cases from observations. In observations, the parallel velocity fluctuations can be obtained from Doppler-shifted spectral lines using velocity centroids, while the plane of sky magnetic field fluctuations can be obtained using polarization direction fluctuations (see \citealt{ch5}). We provide the results corresponding to the analysis of the data corresponding to $k_{\|, G}=0$ Fourier component harmonic. The subscript "G" denotes the measurements taken in the global mean field system of reference. This is the only system of reference available in observational studies.

\begin{figure*}
\centering
\centering
\includegraphics[width=0.98\linewidth]{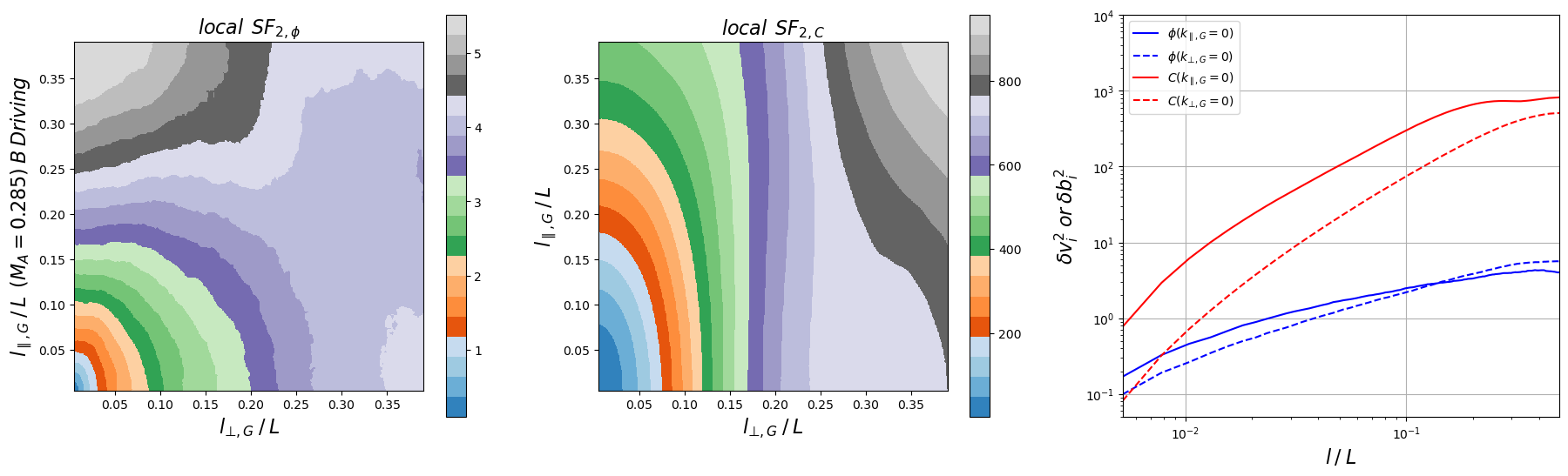}
\\ % Adds a line break between the two images
\includegraphics[width=0.98\linewidth]{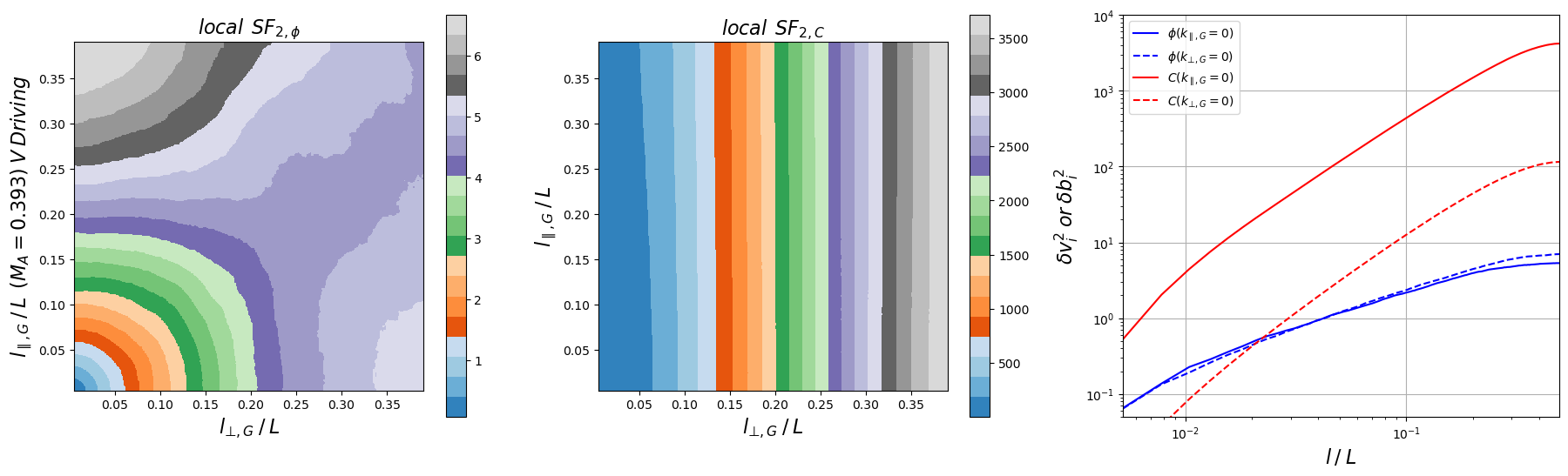}
\caption{Structure functions and contours of equal correlation perpendicular to the line of sight magnetic fluctuations and parallel to the line of sight velocity fluctuations corresponding to $k_\|=0$. 
Left column: Contours of equal correlation for the POS magnetic field fluctuations. Middle column: Contours of equal correlations for velocity centroids. Right column: Structure functions of the POS magnetic field that can be measured through polarimetry and structure functions of velocity centroids available through observations. 
Upper row:  Magnetically driven case.  
Lower row: The same as in the upper panel, but for the velocity-driven case.}
\label{fig:both_figures}
\end{figure*}

A comparison between panels of Figure \ref{fig:both_figures} reveals that for velocity-driven turbulence, the structure-function of velocity parallel to mean magnetic field is significantly larger than the structure-function of velocities measured perpendicular to the magnetic field. On the contrary, the two functions converge at large scales for magnetically driven turbulence. This opens the way for observationally distinguishing between the two cases. We will explore this in a future study.

\section{Comparison with the earlier studies of sub-Alfvenic turbulence}
\label{sec:comp}
\subsection{Velocity driving providing $E_B=E_k$}
\label{sec:EB=EK}

The studies of diffusion induced by sub-Alfvenic turbulence were undertaken in \cite{2021MNRAS.503.1290S}. As elongated boxes and anisotropic driving were employed in the studies, the results are difficult to compare with those in this paper. In particular, the equi-partition between the kinetic and magnetic energies was reported, i.e.
\begin{equation}
    E_{weak, k}=E_{weak, m},
    \label{santos}
\end{equation}
which we attribute to the anisotropic driving that was intended to prevent the excitation of modes with $k_\|=0$.

%The differences cannot be attributed to the use of compressible code in \cite{2021MNRAS.503.1290S}. The sonic Mach number in the simulations was chosen low, and thus the simulations were "nearly incompressible". Moreover, we discuss further the compressible simulations in \cite{2022MNRAS.515.5267B} performed with a setup similar to ours, and their results are consistent with our incompressible simulations. Thus we conclude that the difference in driving and the use of elongated boxes allows sub-Alfvenic turbulence to show different properties. 

The driving in \cite{2021MNRAS.503.1290S} is rather special. We doubt that this type of restricted driving is common in typical astrophyisical environments. More research is required to answer this question. Our study shows that, unlike trans-Alfvenic turbulence, sub-Alfvenic turbulence is more sensitive to the properties of the large scales turbulence driving.

Note that while simulations in \cite{2021MNRAS.503.1290S} confirmed the dependencies of the reconnection diffusion obtained for weak turbulence in \cite{Laz05}, the study did not agree with all the predictions of the classical weak turbulence theory (see \citealt{2007NJPh....9..307N}).

\subsection{Velocity driving providing $E_B/E_k=M_{A,b}$}
\label{sec:strange}

The discussion of the disparity of kinetic and magnetic energies in compressible turbulence can be traced back to \cite{Federrath_2016}, where for $B_0\gg \delta B$, the magnetic energy was presented as 
\begin{equation}
  B^2\approx (B_0+\delta B_\|)^2+\delta B_\bot^2\approx B_0^2+2B_0\delta B_{\|}   
\end{equation}
and the second term, $B_0 \delta B_\|$, was identified with the turbulent part of the magnetic energy\footnote{This involves redefining what turbulent magnetic energy is and disagrees with the established fact that the velocity-driven sub-Alfvenic turbulence has the access of kinetic energy over magnetic energy.} and equated to the kinetic energy
\begin{equation}
    \langle B_0 \delta B_\|\rangle \approx \langle \rho \delta V_L^2 \rangle,
    \label{strange}
\end{equation}
where the $\langle ... \rangle$ denote ensemble averaging and the $V_L$ is the injection speed in compressible MHD simulations. \cite{2022MNRAS.510.6085L}
%https://ui.adsabs.harvard.edu/abs/2022MNRAS.510.6085L/abstract
stated that identifying $B_0 \delta B_\|$ with magnetic energy is unjustified, which agrees with the conclusion in \citealt{2022FrASS...9.3556L}). It is easy to see that Eq. (\ref{strange}) and our Eq. (\ref{2}) differ radically. They differ in the underlying physics, assumptions made during the derivation, and their final form. In particular, our derivation suggests that $|\delta B|$ is the standard deviation of magnetic fluctuation rather than the random variable with zero mean, as it is in Eq. (\ref{strange}).

{ The original derivation of Eq. (\ref{strange}) was reproduced in ST21, but later publications attempted to ensure that the left-hand side of the equation is positively defined, i.e.,  $\langle \delta B\rangle \stackrel{!}{=} \sqrt{\langle\delta B^2\rangle}$ as this was required if $B_0\delta B_\|$ is the dominant part of the fluctuation of magnetic energy. These attempts include the introduction of two different potentials that govern parallel and perpendicular fluctuations of the magnetic field in \cite{2021A&A...656A.118S}. The authors claim that harmonic oscillations of $\delta B$ making the $\langle B_0 \delta B_\|\rangle$ equal to zero are present in the case of incompressible turbulence, but for "compressible strongly magnetized turbulence," the oscillator is not harmonic and the "potential energy of perturbations (here of the magnetic field) is linear, $\delta \epsilon_{p, compressible} \propto \delta B$" rather than to $(\delta B)^2$. However, this conclusion follows from the authors' assumption routed in Eq. (\ref{strange}) rather than an outcome of calculations that were supposed to result in the potential proportional to $\delta B^2$ after the accepted averaging procedure. It is also surprising why $\delta B_\|$ and $\delta B\bot$, according to simulations, stay within a factor of 2 for a wide range of $M_A<1$, despite being governed by potentials of different nature. In contrast, the correspondence of parallel and perpendicular magnetic fluctuations expressed by Eq. (\ref{3}) is natural in our derivation. 

If one first takes squares of both parts of Eq. (\ref{strange}) and then performs averaging, formally, the problem on the left-hand side of the resulting equation stops being zero. This was suggested in \cite{2022MNRAS.515.5267B} (henceforth BX22), where it was proposed to consider the "second moment of the energy balance equations." Such physical quantity had never been used in classical mechanics, and its physical meaning/significance was unclear, as pointed out in \cite{ch5}. In other words, while solving some problems, this approach created new questions. 

Finally, a sophisticated approach involving Lagrangian description of compressible MHD turbulence was suggested in \cite{2023A&A...672L...3S}.
The authors conclude that "for sub-Alfvenic and compressible turbulence, we find that ${ B}_0\delta { B}$ is the leading term in the dynamics, and as a result, the scaling of the velocity and magnetic fluctuations become
$\delta u_l\sim \sqrt{B_0 \delta B_l}$, or equivalently, $M_A\sim \delta u_l/V_A\sim \sqrt{\delta B_l/B_0}$". The former expression corresponds to the relation that follows from Eq. (\ref{strange}), provided that $\delta B$ is positively defined. The authors stress, that in their formalism, "the incompressible limit is approximated when ${ B}_0\delta { B}=0$". This means the Eq. (\ref{M^2}) and related Eq. (\ref{K/B}) should not be true in the incompressible case. This contradicts the results in this paper. At the same time, we suspect that the authors do not account for the fact that to compress a magnetic field, one does not require the fluid to be compressible and magnetic field is being compressed by slow/pseudo-Alfven modes also in the limit of incompressible MHD. Applying the Lagrangian approach to slow modes may be a worthy exercise. 

All of the studies above assume that turbulence compressibility is mandatory for obtaining the relations that we, nevertheless, obtained for incompressible subAlfvenic turbulence. At the same time, none of the earlier studies suspected that the velocity driving induces the effects we reported in the present paper.\footnote{We note parenthetically that in the aforementioned papers, Eq.\ref{strange} is assumed to be true for { any} choice of ensemble volume. In the language of turbulence, this means Eq.\ref{strange} holds for all choices of wavenumber $k$. This is different from our derivation of Eq. (\ref{2}) where we consider the pressure balance only at the injection scale. In the case of compressible supersonic turbulence, the scalings of $\delta \rho$, $\delta v$, and $\delta B$ are different \citep{KowL10} and the energy conservation arguments may not hold for individual modes.} We reiterate that the derivation of Eq. (\ref{2}) suggests that $|\delta B|$ is the standard deviation of magnetic fluctuation rather than the random variable with zero mean, as it is in Eq. (\ref{strange}). This follows from our derivation that is based on the force balance rather than on equating energies. Therefore the justification of Eq. (\ref{2}) does not require any additional hypothesis.}

}

{ For the sake of completeness, below, we mention other differences with the earlier studies of properties of subAlfvenic turbulence.} BX22 stresses the difference in their results from the classical theory of sub-Alfvenic turbulence and conjectures that sub-Alfvenic turbulent fields may not exist for $M_A<2$. Our results contradict the latter hypothesis. This paper predicts and demonstrates through numerical testing a well-defined turbulent scaling.  Our study shows that the turbulent description of random velocities applies to magnetized fluids with a high Reynolds number at any $M_A$. 

In general, the aforementioned studies associate the differences in the outcome of their simulations and the classical MHD theory (e.g., \cite{GS95}) with the effects of compressibility. For instance, they report that fluid compressibility causes the dominance of kinetic energy over magnetic turbulent energy in $M_A<1$ simulations (see \citealt{2021A&A...647A.186S,2021A&A...656A.118S}). On the contrary, our study deals with incompressible turbulence and gets the same relations between $E_k$ and $E_B$.  Formally, the latter does not disprove the former but shows that compressibility cannot be the sole agent responsible for the energy disparity. The striking similarity of the results obtained in the presence and absence of compressibility presents a problem for the earlier theories. We hypothesize that in both cases, the driving is the cause of the disparity. Our hypothesis can be tested by simulating compressible turbulence with magnetic driving. If our hypothesis is correct, we will not see a significant disparity between turbulent magnetic and kinetic energies for magnetically driven turbulence. However, such a study is beyond the scope of the present work, which is focused on the fundamental properties of incompressible subAlfvenic turbulence.

\subsection{Magnetically driven turbulence with $E_B=E_k$}

If chaotic magnetic field motions drive turbulence, kinetic energy does not exceed magnetic energy. Thus, even for $M_A<1$, the turbulence is expected to be in the strong MHD regime at all scales. This was noted in \cite{Eyink13} in relation to turbulence arising from turbulent reconnection (see also \citealt{Laz20}). For strong Alfvenic turbulence,
\begin{equation}
    E_{k}=E_{B}
    \label{total}
\end{equation}
for every scale, starting from the injection scale. An example of how the equipartition is re-emerged is shown in Fig.\ref{fig:B-field_Driving}, where we drive the turbulence mathematically similarly to the kinetic counterpart from the induction equation instead of the momentum equation.

The physical reasoning on why a violation of equipartition is not observed in this case is rather trivial. The cascade time (Eq.\ref{cascade}) given by the small velocity fluctuations transferred from the magnetic field driving is {\it significantly} longer than both timescales (See \S 2.2.2) of weak turbulence. As a result, the system has enough time to develop itself to the equipartition state before hitting the cascade time boundary as outlined in \S 2.2.2.

Turbulence induced by magnetic reconnection is an example of this type of turbulence (See numerical simulations of this front, e.g., \citealt{Kow17,Eyink13, 2020ApJ...901L..22Y}).

This type of sub-Alfvenic turbulence is another example of when the excitation of turbulence affects its properties.
The driving of sub-Alfvenic turbulence in terms of Elsasser variables is frequent in numerical simulations (see \cite{BL19}). This driving is similar to the magnetic driving of turbulence. Instabilities related to the magnetic field are expected to provide magnetic driving (see more from \S \ref{sec{turbdrive}}).  

%\begin{figure}
%\centering
%\includegraphics[width=1\linewidth]{smallVdriving.png}
%    \caption{Energy spectra for both velocity and magnetic field with small scale velocity driving ($k_f=6$) .}
 %   \label{fig:high-k_driving}
%\end{figure}

\section{Properties of magnetic fluctuations}
\label{sec:magfluc}
\subsection{Comparison with theoretical expectations: GS95 and LV99}

The study in GS95 was performed for trans-Alfvenic turbulence. Our results smoothly transfer to this regime as $M_A\rightarrow 1$.

The analytical scaling that involves the Alfven Mach number was introduced in LV99. It was extensively used in subsequent publications to describe different physical processes, e.g., the propagation of cosmic rays (see \cite{YL02,YL03,YL08,Petrosian:2006,XY13,XL16,Krum20,2022ApJ...926...94M,Hucr22} and 
damping of streaming instability by MHD turbulence (see \cite{La16,LX22,2019MNRAS.490.1271H, 2021MNRAS.501.4184H,2022ApJ...928..112B, XL22}).

These results stay intact for the magnetic driving of sub-Alfvenic turbulence. However, for the velocity driving of turbulence, $M_{A,b}$ should be used to describe the dynamics of Alfvenic part of the cascade.

\subsection{Comparison trans-Alfvenic numerical studies: \cite{CL03}}

Numerical decomposition of MHD turbulence into slow, fast and Alfven modes was performed in \cite{CL03} and, with modifications, in the subsequent publications (see \citealt{KowL10,2021ApJ...911...53H,BL19,2020PhRvX..10c1021M,2022MNRAS.512.2111H,leakage,2023MNRAS.520.3857H,2024arXiv240517985P} ). This study uses the original \cite{CL03} approach to explore how energy partition between the modes proceeds in sub-Alfvenic turbulence. We deal only with Alfven and slow/pseudo-Alfven modes for incompressible turbulence. 

\begin{figure}
\centering
\includegraphics[width=1\linewidth]{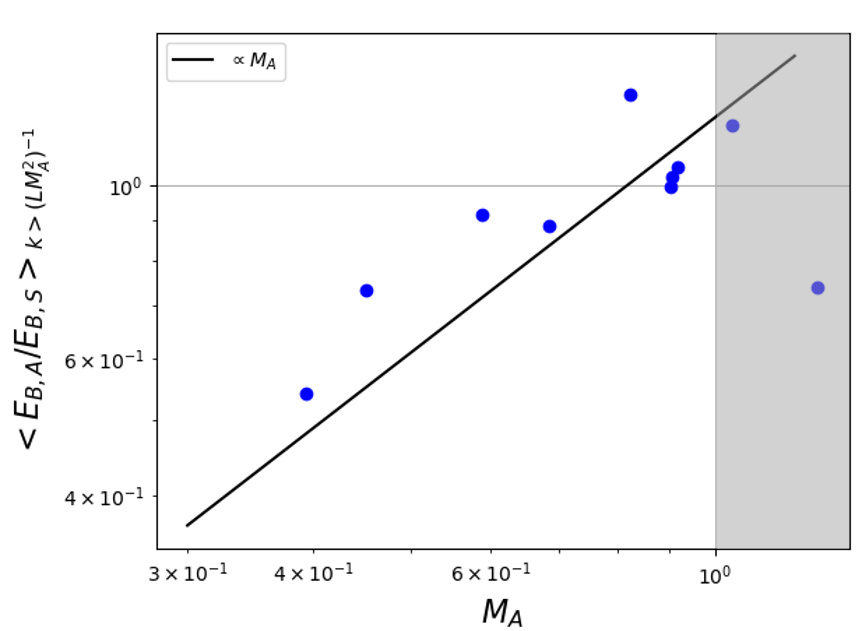}
    \caption{The relation of the mean magnetic spectra ratio of slow and Alfvenic modes for the strong turbulence regime. Grey region represents the region that is Super-Alfvenic.}
    \label{stagnation1}
\end{figure}

We argued above that for the velocity driving of sub-Alfvenic turbulence, the energy of slow modes is expected to exceed the energy of Alfven modes in the regime of strong MHD turbulence. The corresponding testing is presented in Fig. \ref{stagnation1}. This result is consistent with Fig. \ref{fig:sf_temporal}, which shows that the cascading of Alfven and slow modes starts at different scales.

In earlier studies, it was assumed that MHD turbulence properties do not depend on whether the turbulence is driven by magnetic or velocity fluctuations. 

The results of Fig. (\ref{stagnation1}) are different from the results in \cite{CL03}, where for trans-Alfvenic turbulence, the energies of slow and Alfven modes were found to be of the same order. However, our present study shows that for velocity driving of MHD turbulence, this is only true for the case of $M_A\approx 1$. 

The dominance of velocity fluctuations in sub-Alfvenic turbulence entails many astrophysical consequences missed through extrapolating results in \cite{CL03} to the sub-Alfvenic regime. We mention some consequences in \S \ref{sec:disc}.

\subsection{Comparison with sub-Alfvenic compressible numerical simulations: \cite{2020MNRAS.498.1593B}}
\label{sec:beattie20}

\begin{figure}
\centering
\includegraphics[width=1\linewidth]{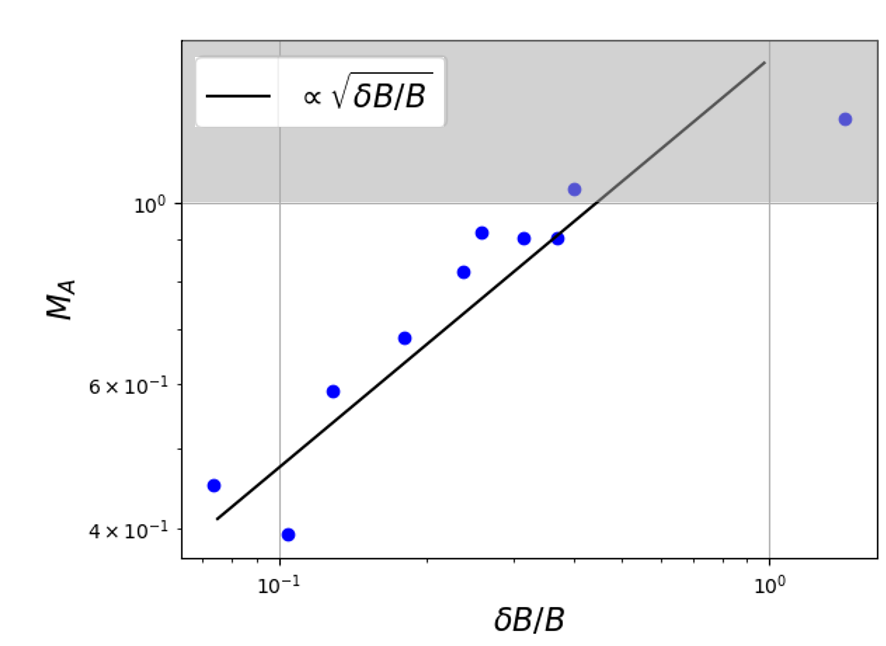}
    \caption{Scaling of $M_{A,b}=\delta B/B$ with the traditional Alfven Mach number $M_A$ for our simulations in Tab.\ref{tab:sim}.The grey region represents the super-Alfvenic domain. }
    \label{stagnation}
\end{figure}

In \S \ref{sec:2.3}, we argued that for the case of incompressible MHD turbulence driven by velocity fluctuations, the normalized magnetic field fluctuation $\delta B/B$ scales as the squared Alfven Mach number $M_A$. This expectation is confirmed by Fig. \ref{stagnation} for our incompressible simulations. 

 In their study of compressible MHD turbulence \cite{2020MNRAS.498.1593B} obtained a similar relation. This correspondence between results obtained for the compressible and incompressible cases {may be} suggestive that for velocity-driven MHD turbulence, the fluid flow along magnetic field lines dominates the velocity dispersion and induces magnetic fluctuations in a way that marginally depends on fluid compressibility. { This hypothesis should be tested by future research. Below we provide a few considerations related to what one may expect due to effects of compressibility. } 

The major difference between compressible and incompressible MHD turbulence is the absence of fast modes in the incompressible case. In addition, incompressible turbulence formally corresponds to the infinitely large ratio of the sound to Alfven speed, i.e., to $\beta=\infty$. For $\beta<1$, the nature of slow modes also changes compared to the high $\beta$ case \citep{CL02_PRL}. The magnetic fluctuations induced by fluid flow along magnetic field lines (see \S \ref{2.2.1}) in low-$\beta$ case correspond to fast modes. Finally, compressible driving employed in \cite{2020MNRAS.498.1593B} changes the ratio between energies of slow, fast, and Alfven modes \citep{2022ApJ...926...94M}. All these effects are expected to alter the relations between the dispersions of magnetic and velocity fluctuations. The fact that Eq. (\ref{M^2}) holds for the compressible settings explored in \cite{2020MNRAS.498.1593B} indicates the sub-dominance of these effects for the velocity-magnetic field relation at the injection scales.

To what extent the results we obtained for an incompressible turbulence case, i.e., for turbulence in media with $\beta=\infty$ and sonic Mach number $M_s$ equal 0, are modified for a wide range astrophysically-relevant $\beta$ and $M_s$ 
requires further studies. One can argue that our considerations in \S \ref{2.2.1} do not directly depend on the media $\beta$. For instance, for sub-Alfvenic turbulence, the low-$\beta$ fluid flow generated by random velocity driving will align with the magnetic field, and the fluctuations of velocity will create pressure fluctuations that will induce $\delta B_\|$ according to Eq. (\ref{2}). This, in turn, will induce the perpendicular fluctuations given by Eq. (\ref{4}) due to a divergence-free constraint on the magnetic field. Similarly, if the fluid motions along magnetic field lines dominate at the driving scale, the velocity dispersions should not be much affected by shocks accompanying turbulence in high sonic Mach number media.

Some astrophysical media is not isothermal. We expect to see a different statistics in this case. For instance, it is reported that in the case of multiphase media, fast mode generations are far more efficient and can arrive up to 50\% of the total turbulence energy due to thermal instability (Beresnyak et al. in prep, see also \citealt{instability,2024arXiv240714199H,2024arXiv241013244H}).

Naturally, the difference between $M_A$ and $M_{A,b}$ disappears for the incompressible MHD turbulence if turbulence is driven, e.g., in terms of Elssasser variables or driven by magnetic fluctuations, as it follows from Fig. \ref{fig:B-field_Driving}. These two Alfven Mach numbers can be somewhat different if magnetic fluctuations drive compressible turbulence. The value and importance of this difference should be established through further research.

Our present study is not limited to establishing the relations between the dispersions of turbulent magnetic fields and velocities. Eq. (\ref{slow_alf}) establishes that the velocity driving in high-$\beta$ fluid induces the dominance of slow modes over the entire turbulence inertial range. %Extrapolating the physics discussed in \S \ref{2.2.1} to the low-$\beta$ case, one can expect a significant increase in the power of the fast modes generated by velocity driving. Testing of these theoretical expectations with compressible MHD simulations is due. 

%\section{Possibility of identifying the regime }

\section{Implications for techniques in estimating magnetic field strength}
\label{sec:imp}
\subsection{Davis-Chandrasekhar-Fermi technique for turbulence with $E_B=E_K$}

Magnetic field strength measurements are important but challenging (see \citealt{Crut10}). Tracing variations of magnetic field $\delta B_\bot$ measured by dust polarization \citep{2015ARA&A..53..501A} 
and combining those with the measures reflecting the velocity fluctuations provides a valuable way of estimating Plane of Sky magnetic field strength in astrophysical settings. The resulting expressions depend on how the fluctuation of $\delta B_\bot$ and the measures of velocity fluctuations are related.

The roots of combining dispersions of magnetic field $\delta \phi$ with Line of Sight (LOS) dispersion of velocity $\delta v$ measured through spectroscopic observations to find the magnetic field strength in molecular clouds can be traced to the pioneering studies \cite{1951PhRv...81..890D} and \cite{ChanFer53}. We refer to the original studies and their subsequent modifications/improvements that employ $\delta \phi$ and $\delta v$ to estimate the magnetic field $B$ as the DCF technique. 

 For small perturbations, the fluctuations of the Plane of Sky (POS) magnetic field direction are  
\begin{equation} 
 \delta \phi\approx \frac{\delta B_\bot}{B}, 
 \label{deltaphi1}
\end{equation} 

Within the DCF technique, magnetic fluctuations arise from Alfven waves with amplitude $\delta B=\sqrt{4\pi \rho} \delta v$. This results in $\delta \phi \approx \sqrt{4\pi \rho} \delta v/B$,
which for angle $\gamma=\pi/2$ between the line of sight and the mean magnetic field provides the well-known DCF expression 
\begin{equation}
    B_\bot\approx f \sqrt{4\pi \rho}\frac{\delta v}{\delta \phi},
    \label{DCF_i}
\end{equation}
where $f$ is a factor that is empirically adjusted to account for the complexity of the realistic settings compared to the DCF model. 

With the advances in our understanding of interstellar medium dynamics (see \citealt{MO07}), we understand that the relations between $\delta \phi$ and $\delta v$ should be obtained via employing MHD turbulence theory.

A closer look at Eq. (\ref{DCF_i}) suggests that the variations of dust polarization reflect the variations of the magnetic field direction perpendicular to the line of sight, i.e., $B_\bot$, while the Doppler broadening is affected by the fluctuations of the line of sight components of turbulence velocities, i.e., $\delta v_\|$.\footnote{To justify DCF approach, one should assume that the magnetic and velocity fluctuations arise from the same magnetized motions. Assuming turbulence isotropy, the technique relates the polarization measures perpendicular to the line of sight magnetic field fluctuations, i.e., $\delta B_\bot$, with the Doppler broadening providing the line of sight fluctuations $\delta v_\|$. } In Alfvenic turbulence, the velocity fluctuations are perpendicular to that of the magnetic field, which induces the dependence of factor $f$ on the angle $\gamma$ between the 3-dimensional direction of the magnetic field and the line of sight. Accounting for the existence of slow and fast modes in MHD turbulence (GS95, \citealt{LG01,CL03}), makes $f$ dependent on $\gamma$ and the relative contribution of fundamental modes into MHD turbulence. For the magnetically driven turbulence, one can write:
\begin{equation}
   B_\bot\approx f_1(\gamma) \sqrt{4\pi \rho}\frac{\delta v_\|}{\delta \phi},
    \label{DCF}
\end{equation}
the velocity dispersion is measured parallel to the line of sight, and the angle dispersion is measured perpendicular to the line of sight.

Historically, the dependence of $f_1(\gamma)$ on $\gamma$ is not given enough attention with most of the numerical studies focused on $\gamma=\pi/2$ case. This is an omission, which is crucial to correct while studying anisotropic sub-Alfvenic turbulence, as this paper demonstrates. 

A simplification comes from the fact that  $\delta \phi$ is dominated by the Alfvenic modes (LV99). However, other modes contribute to $\delta v_\|$. This makes the functional dependence of $f (\gamma)$ non-trivial even in the most studied case of B-driving of turbulence (See \citealt{leakage,2023MNRAS.524.6102M} for a method of obtaining $\gamma$ and \cite{2024ApJ...965...65M} for an application.).

\subsection{DCF for turbulence with equi-partition violated}

We note that Eq. (\ref{DCF}) relies on the general equi-partition of magnetic and kinetic turbulent energies, which should not be taken for granted according to the present paper. 

The earlier researchers touched upon the importance of turbulence compressibility (see \citealt{2003LNP...614..252O}) for DCF. 
Our study shows that the outcome depends much more on the turbulence driving.

As we discussed in earlier, the equi-partition between turbulent magnetic and kinetic energies is strongly violated at $M_A\ll 1$. Thus, Eq. (\ref{DCF}) must be modified to account for the modification of turbulence. 

 In the case of velocity driving, the fluctuation of the magnetic field direction given by Eq. (\ref{deltaphi1}) is related to $M_A$ via Eq. (\ref{M^2}). 
Combining this with the definition of the "velocity Alfven Mach number" given by Eq. (\ref{Alfven}), one can easily get the
modified DCF expression:
\begin{equation}
    B_{\bot}\approx f_2(\gamma)\sqrt{4\pi \rho} \frac{\delta v}{(\delta \phi)^{1/2}},
    \label{new}
\end{equation}
where $f_2(\gamma)$ is another factor that depends on the direction of the magnetic field and the angle between the mean magnetic field and the line of sight. In general, $f_2(\gamma)$ is different from $f_1(\gamma)$ and both factors should be determined through dedicated studies similar to one in \cite{ch5}.

Eq.(\ref{new}) was first suggested in ST21 as the substitute for the DCF equation for compressible MHD turbulence, as opposed to the incompressible MHD turbulence that we deal with in this paper. Their approach  was termed in \cite{2022MNRAS.510.6085L}
the "parallel-$\delta B$ version of DCF" as the derivation was based on Eq. (\ref{strange}) that involves only the parallel component of magnetic fluctuation $\delta B_\|$ (see also \cite{2023A&A...672L...3S}). As discussed earlier,  Eq. (\ref{strange}) includes averaging, which makes the left-hand sight of the equation zero by the very definition of the magnetic field fluctuation.  Apart from accepting Eq. (\ref{strange}), the parallel-$\delta B$ version of DCF required another important assumption. {\it According to Eq. (\ref{deltaphi1}), DCF study requires the knowledge of the perpendicular component of the magnetic field.}  
Therefore, an additional assumption of the equality of parallel and perpendicular fluctuations, i.e., $\delta B_\bot\approx \delta B_\|$, was required to make "parallel-$\delta B$ version of DCF" applicable to magnetic field studies. Based on empirical study, this additional assumption justifying the approach was added by \cite{2021A&A...656A.118S}.  More recently, theoretical attempts were made to explain the relations that appealed to turbulence compressibility \citep{2023A&A...673A..76S}.  In contrast, in \S \ref{2.2.1}, we derived the relation $\delta B_\|\approx \delta B_\bot$ from the condition that the magnetic field is divergence free. We did not appeal to any compressibility effects for doing this. We also successfully tested our prediction numerically for incompressible MHD turbulence.

The difficulties with understanding the physical picture behind the ST21 approach were discussed in \cite{2022MNRAS.510.6085L} and in \cite{ch5}. The latter paper, however, stated that there was "no reason to question the ST21 equation as a consequence of an empirically established fact".  The fact that a similar equation was proven to hold for compressible simulations (see ST21, \citealt{2021A&A...656A.118S}) and our incompressible simulations may suggest that the effects of velocity driving are the dominant effect.

Within this study, we do not discuss the $\gamma$-dependence of $f_2(\gamma)$. Empirically for $\gamma=\pi/2$ Eq. (\ref{new}), the coefficient or $f(\pi/2)=2/\sqrt{2}$ was proposed by ST21. Based on the analysis in \cite{ch5}, we expect this numerical value to vary depending on the actual properties of the cascade, i.e., its composition in terms of Alfven, slow, and fast modes.

In this paper, we get the expression given by Eq. (\ref{new}) based on considering a velocity-driven sub-Alfvenic turbulence. Theoretical considerations in GS95 suggested that the properties of the Alfvenic cascade are marginally affected by compressibility. For the case of trans-Alfvenic turbulence, this was numerically demonstrated that the energy exchange of Alfven and fast/show modes is marginal (\citealt{CL03}, see also \citealt{CL05}).
 % https://ui.adsabs.harvard.edu/abs/2005ThCFD..19..127C/abstract
 This is reasonable to expect that this is also true for sub-Alfvenic turbulence. The presence of the fast modes only marginally changes the variations of magnetic field direction (LV99).  
 Therefore, expect that the fluctuations in magnetic field direction are determined by Alfvenic turbulent cascade and the dependence on $\delta \phi$ that we revealed for our incompressible simulations to be valid for compressible turbulence of molecular clouds. The comparison of our incompressible results with the available compressible studies that we discuss in \S 6.3 supports this conclusion. We stress the importance of the function $f_2 (\gamma)$ that we will define for different regimes elsewhere.

As the regime of weak turbulence is not robust, and it requires further studies to determine which type of turbulence driving is relevant to the astrophysical settings. For instance, the driving in \cite{2021MNRAS.503.1290S} induces turbulence for which the magnetic field strength should be studied using Eq. (\ref{DCF}). 

Both Eq. (\ref{DCF}) and (\ref{new}) assume that the dispersions of magnetic field and velocities are measured over the injection scale $L$. This is not always applicable. In many cases, the velocity dispersion $\delta v$ is measured over the full depth of the cloud $L_c$, while the dispersion of magnetic field angles is measured over the available patch $L_p\times L_p$, where $L_p<L_c$. For magnetically driven turbulence, $\delta \phi$ is artificially decreased, which biases the magnetic field obtained with Eq. (\ref{DCF}) towards higher values. This effect is less pronounced for velocity-driven turbulence, as shown in Fig. \ref{fig:sf_temporal}. 

The difference in the variations of the positional angle with the change of the averaging block can be used to distinguish the two regimes of turbulence driving. This information is significant by itself. In terms of the DCF technique, it determines whether Eq. (\ref{DCF}) or Eq. (\ref{new}) should be used for obtaining the magnetic field strength.

%Consider a cloud with the scale $L_c$ that is part of a big turbulent cascade, but $L_c<L$. This model is consistent with the results in Yuen et al. (2022), where, based on observations of different lines, a universal cascade of turbulence was demonstrated in the direction of the Taurus cloud. 

\subsection{Other approaches to DCF}
\subsubsection{Universal expression for DCF: MM2 approach}

In \cite{ch5}, modifying the DCF equations to use the "velocity Alfven Mach number" $M_A$ instead of the $\delta \phi$ was suggested. The motivation for this was that $M_A$ could be measured not only through polarimetry but with other techniques, for instance, using velocity gradients \citep{Laz18}, Tsallis statistics \citep{EL10}, 
%https://ui.adsabs.harvard.edu/abs/2010ApJ...710..125E/abstract
velocity anisotropy \citep{EL11,LP12,BurL14,2016MNRAS.461.1227K}
%https://ui.adsabs.harvard.edu/abs/2011ApJ...740..117E/abstract
%https://ui.adsabs.harvard.edu/abs/2012ApJ...747....5L/abstract
%https://ui.adsabs.harvard.edu/abs/2014ApJ...790..130B/abstract
%https://ui.adsabs.harvard.edu/abs/2016MNRAS.461.1227K/abstract
. It is important that some of the techniques measuring $M_A$ can obtain 3D information using velocity channel maps, e.g. 
\citep{2015ApJ...814...77E,Laz18}
%https://ui.adsabs.harvard.edu/abs/2015ApJ...814...77E/abstract
 or the synchrotron polarization tomography \citep{2018ApJ...865...59L},
%https://ui.adsabs.harvard.edu/abs/2018ApJ...865...59L/abstract
providing better insight into media magnetization. 

As shown below, this approach provides a universal form for the DCF that is valid for both velocity and magnetic field-driven turbulence.
Eq. (\ref{DCF}) and Eq. (\ref{new}) can be presented in the universal form. 
Indeed, or velocity-driven case, $\delta \phi$ given by Eq. (\ref{deltaphi1}) scales as $M_A^2$ according to Eq. (\ref{M^2}). Thus, instead of Eq. (\ref{new}) one can write
\begin{equation}
B_\bot \approx f (\gamma) \sqrt{4\pi \rho} \frac{\delta v}{M_A}
\label{univ1}
\end{equation}
Incidentally, the magnetically driven case, $\delta \phi \approx M_A$ and Eq. (\ref{DCF}) also transfers into Eq. (\ref{univ1}). The functions $f(\gamma)$ may be different, and the difference should be determined by future studies.

One can take the next step and employ sonic Mach number $M_s\approx V_L/c_s$, where $c_s$ is sound velocity instead of $\delta v$ in Eq. (\ref{univ1}). This number can be measured by several techniques described in the literature \citep{Burk09,EL10,Burk12,Burk16}.

%https://ui.adsabs.harvard.edu/abs/2009ApJ...693..250B/abstract
%https://ui.adsabs.harvard.edu/abs/2010ApJ...710..125E/abstract
%https://ui.adsabs.harvard.edu/abs/2012ApJ...749..145B/abstract
%https://ui.adsabs.harvard.edu/abs/2016ApJ...827...26B/abstract

Then, one gets the expression of the magnetic field obtained in \cite{ch9}:
\begin{equation}
    B_\bot\approx f_3 (\gamma) \sqrt{4\pi \rho} c_s \frac{M_s}{M_A}
    \label{univ2}
\end{equation}
which universally applies irrespectively of the turbulence driving. This technique was termed in \cite{ch9} MM2 to reflects that it uses two Mach numbers, the Alfven and sonic one.

\subsubsection{Broad view of MM2 approach}

The advantage of this technique, termed MM2 in \cite{ch9}, as it used two Mach numbers, is that it applies to various data sets for which $M_A$ and $M_s$ can be measured. This can be done 
 by analyzing spectroscopic channel maps, 
 where measuring the dispersion of parallel velocity is impossible. Nevertheless, both $M_A$
 and $M_s$ can be obtained using Velocity Channel Gradients (VChG) \citep{Laz18,2020ApJ...898...65Y}.
%https://ui.adsabs.harvard.edu/abs/2020ApJ...898...65Y/abstract
The MM2 technique has already been applied to several spectroscopic data sets, e.g. 
\cite{2021ApJ...912....2H}.
%https://ui.adsabs.harvard.edu/abs/2021ApJ...912....2H/abstract

If the statistics of the fluctuations are related to $M_A$ and $M_s$ by using numerical simulations that are approximate the actual astrophysical settings, one can use the universal Eq. (\ref{univ1}) and (\ref{univ2}) without worrying whether the turbulence is magnetically or velocity driven. 

As we discussed above (see Figure \ref{fig:both_figures}) one can determine from observations whether the turbulence is velocity or magnetic field-driven. Therefore, if the technique provides $M_{A,b}$, it can be transferred to $M_A$ using Eq. (\ref{M^2}). This looks promising for obtaining magnetic field strength for various astrophysical settings. This approach deserves a dedicated study.

It is important to stress that MM2 is not just another way of rewriting the DCF formulae. It opens new horizons for new types of observational data in order to obtain magnetic field strength. For instance, the MM2 technique can also be applied to synchrotron intensity and column density maps.

Machine Learning (ML) also provides a new promising way of $M_A$ and $M_s$ (see \cite{2024MNRAS.52711240H}). Machine learning can be used to determine whether turbulence is driven by magnetic or velocity fluctuations.

\subsubsection{Filtering of the data within the DCF, comparison with DMA}

Our study demonstrated that the difference in $M_{A}$ and $M_{A,b}$ arises from fluid flow along magnetic field lines. This suggests that if the velocity contributions arising from such motions are removed, the $M_A$ and $M_{A,b}$ will coincide. This also means one can use the traditional DCF given by Eq. (\ref{DCF}) without additional concerns about the nature of turbulence driving. 

The motions along magnetic field lines for the observer are seen as motions along the mean field with the perpendicular dispersion $M_{A,b}$. By filtering out the 2D spacial frequencies $K_\|=0$ and $K_\bot$ in the range of $[-M_{A,b}, M_{A,b}]$ one can remove the spacial frequencies that are affected by the fluid parallel flows. A detailed study of this approach will be done elsewhere.

Finally, we briefly compare the DCF and a more recently suggested Differential Measure Analysis (DMA) (\cite{ch5}) approach. Within DMA not the dispersions of magnetic field directions and linewidths but locally measured structure functions of velocity centroids and magnetic field directions are employed. The DMA approach provides a lot of advantages, e.g., it can increase accuracy and provide detailed distributions of magnetic field. However, DMA requires more information about the turbulence properties. We treat DMA and DCF as complementary tools.

\section{Discussion}
\label{sec:disc}

\subsection{Sub-Alfvenic turbulence: Comparison with selected earlier studies}

\subsubsection{Incompressible MHD turbulence studies}

Our paper demonstrates that the difference in how turbulence is driven significantly changes the properties of MHD turbulence. 
The major finding of the present paper is the introduction of the magnetic Mach number $M_{A,b}$ that coincides with $M_A$ for magnetically driven turbulence and is equal to $M_A^2$ for velocity-driven turbulence. $M_{A,b}$ enters the LV99 expressions describing the properties of sub-Alfvenic MHD turbulence at small scales and the expressions that employ the LV99 formalism. Those include expressions for turbulent damping of Alfven waves, heat and matter transfer, removal of magnetic fields from star-forming clouds, superdiffusion of cosmic rays, etc. 

MHD turbulence is a subject with an extended history of research (see, e.g., book by \citealt{BL19}). 
The different regimes of MHD turbulence have been studied theoretically (e.g., \citealt{2000JPlPh..63..447G,2007NJPh....9..307N,2023JPlPh..89b9005G,4DFFT_p1}), numerically \citep{2016PhRvL.116j5002M,2021MNRAS.503.1290S}, and observationally \citep{2024NatAs.tmp...71Z,2024ApJ...962...89Z}.

Weak Alfvenic turbulence is different from its strong counterpart that takes place at scales less than $LM_{A,b}^2$ (LV99, \citealt{2016PhRvL.116j5002M}). In \S 2, we stressed the importance of the perpendicular scale $LM_{A,b}$, which defines the scale at which the magnetic field fluctuations are decorrelated unless enforced by numerical schemes. Our paper demonstrates that turbulence driving profoundly affects the properties of MHD turbulence. The modification is not limited to the weak turbulence range but carries over by changing the ratio of energy in Alfven and slow modes for the strong MHD turbulence regime (see Eq. (\ref{slow_alf})). 

%Why didn't previous numerical simulations on weak turbulence observe this scaling? The "decorrelated case" appearance is highly restricted to Alfvenic turbulence since only Alfvenic turbulence is developed perpendicular to its propagation direction. For the case of compressible turbulence, the cascade is more complex \cite{CL03} since additional transition scales arising from slow modes are present.  In other words, to observe the effect of the decorrelated magnetic field amplitudes, one needs to have the following conditions: (i) Pure Alfvenic turbulence, (ii) having a perpendicular length longer than $L_\parallel M_A$, and (iii) velocity driving of turbulence.

Alfvenic turbulence was studied in many numerical papers, e.g., 
\cite{MG01,2001PhPl....8.2673M,2012MNRAS.422.3495B,2012PhPl...19e5902M,2015MNRAS.449L..77M,2016PhRvL.116j5002M}. Some of them, e.g. \cite{MG01,2016PhRvL.116j5002M} drive both velocity and magnetic field simultaneously while using Elsasser variables (\cite{1950PhRv...79..183E}) via white the noise method (e.g. \citealt{2001PhPl....8.2673M,2012MNRAS.422.3495B,2012PhPl...19e5902M,2015MNRAS.449L..77M}).
This driving implies that the velocity and the magnetic field are driven coherently. As driving by velocity or magnetic field does not have advantages, the theoretical elegance of the Elsasser equation approach \citeauthor{1950PhRv...79..183E} is favored for incompressible studies compared to using Eq. (\ref{eq:eom}). The Elsasser driving conditions ensure the equality of magnetic and kinetic energies. However, such simulations cannot answer what happens in the case of astrophysical driving that can be induced through random velocities.

Our study reveals that the isotropic velocity driving of incompressible sub-Alfvenic turbulence induces a new effect of flows along magnetic field lines. At the injection scales, these flows absorb most of the injected energy, as the dynamically dominant magnetic field strongly resists the driving perpendicular to magnetic field lines. The variations of the parallel velocities $V_L$ induce fluid pressure variations $\delta P$ according to the Bernoulli equation, {and this results in Eq. (\ref{2}) which relates the standard deviation of the parallel magnetic field fluctuation with the fluctuation in the dynamical pressure of the fluid $1/2\rho V_L^2$.} Unlike the magnetic or Elsasser variable driving, for $\delta B_\|< B$, this induces the disparity of kinetic and magnetic energies in turbulence (see Eq. \ref{K/B}). The magnetic field is divergence-free, which ensures the approximate equality of the magnetic fluctuations perpendicular and parallel to the mean field, i.e., $\delta B_\|\approx \delta B_\bot$. The dependence of subAlfenic turbulence properties on the nature of turbulence driving is one of our major findings in this paper. 

The closest to our approach is \cite{CLV_incomp}, where the authors solve Eq.\ref{eq:eom} similarly to our current study. Their set-up provides $M_A \approx 1$, meaning that the trans-Alfvenic turbulence is studied. The condition outlined in \S 2.2 converts to the perpendicular structure function for the magnetic field at large scale $SF_B \propto l_\perp^0$, while that for velocity $SF_v \propto l_\perp^1$. Fig.3 of \cite{CLV_incomp} shows a difference in structure-function at the largest scale, consistent with our present study. 

The accepted description of incompressible MHD turbulence assumes that kinetic and magnetic energies are equal for Alfvenic modes and that the energies of Alfven and slow modes for isotropic incompressible driving are also similar. Our study reports the violation of these two common assumptions if sub-Alfvenic turbulence is generated by isotropic velocity driving. This changes the starting point for accounting for compressibility effects in MHD turbulence.

\subsubsection{Relation to compressible MHD turbulence studies}

Our findings shed new light on the earlier studies of compressible MHD turbulence, which reported the dominance of kinetic over magnetic energies in \cite{2022MNRAS.515.5267B}.  This and the related studies in \cite{2023A&A...672L...3S} attributed this disparity to the compressibility of MHD turbulence they dealt with, as opposed to the classical MHD turbulence theory, e.g., GS95 theory, formulated in incompressible limit. We demonstrate that the same disparity is present in incompressible turbulence. {This does not prove that the theories explaining the effect through compressibility are wrong. However, this, at least, testify that compressibily is not the sole reason for the disparity. For us,} this is suggestive that the energy disparity is a part of the fundamental properties of velocity-driven MHD turbulence already present in the incompressible limit.  {We predict that the magnetically-driven compressible turbulence will not show the same significant difference between the velocity and magnetic turbulent energies.}

{It is well-known that incompressible MHD turbulence is a reasonable approximation for realistic subsonic compressible turbulence in high-beta media (see GS95). For subAlfvenic turbulence in high-beta media, the subsonic requirement is automatically fulfilled. Therefore, the model of turbulence we quantified applies to a wide range of astrophysical high-beta plasmas, e.g., plasma in the warm and hot galactic gas, intercluster gas. Effects of compressibility in such nearly-incompressible gas are marginal, which makes the alternative explanations based on turbulence compressibility unlikely.\footnote{Indeed, appealing to velocity driving in incompressible turbulence, we get the disparity of kinetic and magnetic energies given by Eq. (\ref{K/B}). It is not conceivable that marginal compressibility in high-beta media not only overrides this effect, but imposes the same relation given by Eq. (\ref{K/B}).} 

Potentially, the effects of compressibility may take over in low-beta media. For us, it seems surprising that very different physics will provide the same result given by Eq. (\ref{K/B}). The testing with magnetic driving of turbulence in low-beta media that we proposed can definitively answer whether theories that explained the disparity appealing to the effects of compressibility are correct. 

At present, one may argue that Occam's razor principle suggests that there is likely to be one universal process makes Eq. (\ref{K/B}) correct both in high and low beta media. We argued above that the compressibility explanation does not work in high beta media. An analogy in the history of turbulence research was an empirical discovery of the rapid decay of turbulence in low-beta media. Initially, it was universally accepted to be due to coupling of Alfvenic and compressible modes. At the same time, the decay of turbulence over one eddy turnover was the theoretical prediction of incompressible MHD turbulence (GS95, LV99), which was confirmed numerically (Cho \& Vishniac 2000). Later numerical research confirmed that in both low and high beta media, the decay happens due to the fundamental non-linearity of Alfvenic modes, which is marginally influenced by fluid compressibility (Cho \& Lazarian 2002). We hope that new simulations will soon test the application of Occam's razor principle to the case at hand. 

In short, in terms of the disparity of kinetic and magnetic energies in subAlfvenic simulations our scope of claims is limited to 
\begin{itemize}
    \item We disprove that the presence of kinetic and magnetic energy disparity requires that the physical system is compressible.
    \item We demonstrate that in incompressible fluid, isotropic velocity driving of turbulence creates disparity, while magnetic driving does not induce such a disparity.
    \item We suggest a {\it hypothesis} that the disparity in a compressible case can also result from the isotropic velocity driving and provide a way to test our hypothesis.  We invite the community to test this hypothesis.
\end{itemize}

At the same time, we stress that the disparity between kinetic and magnetic energies given by Eq. (\ref{K/B}) in incompressible subAlfvenic turbulence is not the only valuable result of the paper. We also provided predictions and testing for the spectra of velocities and magnetic field, spectra slow and Alfven modes applicable to high-beta media. Our study poses an  interesting problem of testing to what extend the corresponding relations are different for low-beta compressible media. This can be studied by analyzing the existing subAlfvenic compressible simulations.
}

\subsubsection{Modified DCF: the importance of the way turbulence is driven}
 As our study deals with incompressible turbulence, we do not discuss theories suggested to justify modifying the DCF due to effects of compressibility (see \cite{2021A&A...656A.118S, 2023A&A...672L...3S}). Instead, we address the modifications the DCF technique required for {\it incompressible} turbulence when turbulence is velocity-driven at the injection scale. The numerical evidence in the aforementioned studies testifies that our DCF modification for the case of velocity driving also approximately holds for the compressible case.  While a quantitative study of compressibility's role in improving the accuracy of the DCF approach is beyond the scope of the present paper, below, we provide a few considerations that may justify why compressibility effects may not significantly change the DCF approach. 

From what we know about the relation of compressible and incompressible MHD (see \citealt{BL19}), we expect compressibility to modify the energy partition between fast mode on the one hand and slow and Alfven modes on the other. However, based on \cite{CL03}, we expect the energy partition between the slow and Alfven modes to be similar to the one reported in this paper. As Alfven modes dominate magnetic field wandering (LV99), we expect Eq. (\ref{M^2}) to hold approximately. We also expect that the flows corresponding to $k_\|=0$ will be similar in compressible and incompressible cases. In the case of velocity driving, we may expect the velocity fluctuations for $M_{A,b}<1$ to be dominated by flows along magnetic field lines rather than fast modes. This can explain why Eq. (\ref{new}) derived for the incompressible MHD turbulence in the presence of isotropic velocity driving can be valid for compressible media, which is in agreement with numerical simulations in \cite{2021A&A...656A.118S, 2023A&A...672L...3S}. We are unaware of DCF validity studies using magnetic driving at the injection scale, so we have nothing to compare Eq. (\ref{DCF}) for the compressible case. 
Further studies employing the separation of modes following the procedures in \cite{CL03} and \cite{KowL10} will be done elsewhere to test these expectations. 

\subsection{Impact on understanding of astrophysical processes: selected implications}

\subsubsection{Improved analytical description of MHD turbulence}

Below, we primarily discuss how our present results modify the conclusions reached within the earlier studies of MHD turbulence that did not consider the effects arising from velocity driving. Naturally, the list of implications is very extensive and deserves a separate study.

The transition scale from weak to strong turbulence (see \S \ref{sec:subAlfven}) and the corresponding small-scale scaling of strong MHD turbulence are functions of $M_{A,b}$ (LV99). For the velocity-driven turbulence $M_{A,b}$ must be related to the traditional Alfven Mach number $M_A$ via Eq. (\ref{M^2}). 
Thus, in the case of velocity driving of sub-Alfenic turbulence, only a fraction of total velocity dispersion at the injection scale $L$ can be associated with the driver of the Alfvenic cascade.

In particular, using $M_{A,b}$ is important for describing statistics of MHD turbulence that can be obtained from observations \citep{LP12,2017MNRAS.464.3617K,2017MNRAS.470.3103K,2016MNRAS.461.1227K} and applying the corresponding tools to explaining the observations. In particular,
\cite{2018MNRAS.478..530K} found that the Plank result of B/E power ratio is consistent with interstellar media being sub-Alfvenic. If turbulence in ISM is velocity-driven, then the Alfven Mach number in \cite{2018MNRAS.478..530K} should be identified with $M_{A,b}$. The upper limit that was found in \cite{2018MNRAS.478..530K}, namely,
$M_{A,b}<0.5$ satisfies the observational data. This limit translates, according to Eq. (\ref{M^2}), into $M_A<0.7$, which is less restrictive in terms of media magnetization.

\subsubsection{Diffusion and acceleration of cosmic rays}

The analytical theory of MHD turbulence is critical for describing the propagation and acceleration of cosmic rays (CRs, see \citealt{Schlickeiser02}).
Sub-Alfvenic turbulence is typical for diffuse ISM and galactic halo, where CRs propagate \cite{Ginzburg}.
In addition, turbulence can be important for the stochastic acceleration of energetic particles (see \cite{Ptuskin88,Blas00, Brunetti_Laz}). 

The most trivial consequence of our paper is that in the CRs studies, $M_{A,b}$ should be employed. This requires distinguishing the cases of velocity and magnetic field-driven turbulence.

The amplitude and anisotropy of the magnetic fluctuations critically affect both the CRs acceleration and propagation through resonant scattering (see \cite{Chanmc00,YL02,YL03,XY13,Hu22sh,2022ApJ...926...94M}).
Our results show that for turbulence's isotropic velocity driving, the magnetic fluctuations' spectrum scales as $k^{-1}$ for scales from $LM_A$ to $L$. This has significant consequences for the magnetic field wandering that determines perpendicular to mean magnetic field superdiffusion of CRs \citep{LY14}.

It is also significant that Alfvenic turbulence is an important factor in suppressing the streaming instability that regulates the transport of CRs \citep{YL02,FG04,Laz16}. 
Incidentally, this effect is also essential for driving galactic winds (see \cite{LX22}).

Another funding in this paper is that the slow mode magnetic fluctuations do not change their amplitude in the $[LM_{A,b}, L]$ range for the velocity driving in high $\beta$ medium. These fluctuations dominate the effect of magnetic mirror diffusion \citep{XL20,2021ApJ...923...53L} and acceleration \cite{2023ApJ...956...63L}.

Transient Time Damping (TTD) acceleration has a vital effect on CR propagation and acceleration (see \cite{1998APh.....9...79R,2019MNRAS.490.1271H,2007HiA....14...97B}.
Magnetic compressions associated with slow modes can dominate the process (see \cite{XLb18}).
Our study shows that the amplitude of such compressions may be $\sim M_A^{-1}$ larger than the Alfven ones (see Fig. (\ref{stagnation1}). This factor should increase the importance of slow modes in TTD acceleration for high-$\beta$ medium. 

We do not dwell upon the quantitative consequences of these effects. The improved understanding of sub-Alfvenic turbulence attained in the present study deeply impacts CR physics. The quantitative discussion of CRs' propagation and acceleration will be provided elsewhere.

\subsubsection{Evolution of interstellar medium and star formation}

Magnetized turbulence is critical in star formation \citep{MO07}. Thus, accounting for the disparity of kinetic and magnetic turbulent energies is vital for understanding star formation processes. More energy associated with fluid motions along magnetic field lines tends to produce a more prominent turbulent concentration of matter. This is important for the formation and evolution of molecular clouds. In particular, this can contribute to forming density structures perpendicular to the magnetic field \citep{XJL19,2020MNRAS.492..668B}. 

At the same time, according to the model of the formation of filaments parallel to the magnetic field (see \citealt{Xupar19}), more energy in slow modes reported in our study results in a higher contrast of such filaments. For instance, this can provide the compression necessary to explain the higher rate of CO formation within the striation observed in Polaris Flare \citep{2023A&A...673A..76S}.

The relative increase of the slow mode amplitude can be one of the causes of the appearance of the Small Ionized and Neutral Structures (SINS) observed in the interstellar medium and discussed in Heiles (1988).
%https://ui.adsabs.harvard.edu/abs/1989ApJ...336..808H/exportcitation
The alternative suggestions attempting to explain these mysterious structures include the current sheets in fully ionized plasmas \citep{2006ApJ...640L.159G} and a viscosity-damped regime of MHD turbulence in partially ionized gas \citep{CLV03,2007ASPC..365..324L}. Both suggestions have shortcomings; thus, accounting for enhancing the relative amplitude of pseudo-Alfven/slow modes we reported in the paper is important. One can argue that the corresponding magnetic fluctuations cause the enhancements of ionized gas, which can alleviate solving the SINs puzzle. However, to what extent this can be true requires more studies of compressible MHD turbulence in partially ionized gas (see \cite{2024MNRAS.527.3945H}).
%https://ui.adsabs.harvard.edu/abs/2024MNRAS.527.3945H/exportcitation

The process of magnetic field diffusion induced by turbulent reconnection, i.e., the reconnection diffusion \citep{Laz14r,Lazarian06,Laz05,Sant10}
is an essential process for the removal of magnetic fields from molecular clouds. The predictions for the rate of the reconnection diffusion in sub-Alfvenic case were confirmed in \cite{2021MNRAS.503.1290S} for the case that corresponds to $E_k=E_B$ (see \S \ref{sec:EB=EK}). For velocity driving of turbulence, the analytical dependences for the reconnection diffusion rate on $M_A$ should be reconsidered.

We note that numerous earlier numerical studies dealing with interstellar turbulence \citep{2008ApJ...688L..79FZ,2010ApJ...715.1302V,2011MNRAS.411...65B,2022MNRAS.515.5267B} reported the importance of the driving of fluid motions along the magnetic field. Our present study relates these flows with the velocity driving of turbulence and quantifies the resulting statistics. 

\subsection{Drivers of galactic turbulence}
\label{sec{turbdrive}}

The nature of galactic turbulence is a hotly debated subject. Stellar winds, supernovae explosions, and Magneto-rotational instability are examples of turbulent drivers proposed in the literature (see \cite{MO07}). On the contrary, some authors advocate turbulence generation through gravitational instability (see \cite{2017MNRAS.467.1313V}). Our study demonstrates the significant role that driving can have on the properties of turbulence.  

Observational studies of galactic turbulence \citep{Armstrong95,CheL10,Xu_Zhang_turb,Yuen22}
suggest the existence of a universal cascade that starts at the scale comparable with the galactic scale height and proceeds to the dissipation scale over many orders of scales. This is consistent with driving on scales significantly exceeding the scale of molecular clouds. Turbulence driving at the scale height of the galactic disk provides a possible explanation of the observed turbulence properties. Expanding supernovae superbubbles can be a possible driver along with gravitational perturbations responsible for the emergence of the Radcliff wave \citep{2020Natur.578..237A}.  

We expect that most of the instabilities are induced by either stellar evolution feedback or gravitational instabilities to induce turbulence through velocity driving. The consequences that follow from our study arise when the turbulence is driven isotropically by sub-Alfvenic driving. It is unclear to what extent this applies to our galaxy or other spiral galaxies. 

Within the picture of the global turbulence cascade, it is not reliable to evaluate the nature of turbulence driving by detecting sub-Alfvenic turbulence over a small patch of the sky. The observed anisotropy of turbulence may result from the measurements over scales significantly smaller than the driving scale. At the same time, as discussed in the paper, the measurements of magnetic fluctuations across scales and their comparison with velocity fluctuations can answer whether the sub-Alfvenic turbulence was velocity-driven. Thus, the peculiar features of sub-Alfvenic turbulence we revealed in the present study should be relevant in the galactic context. 

Like most papers dealing with MHD turbulence, our paper assumes an isotropic turbulence driving. This assumption is not guaranteed to be true in realistic astrophysical settings. Moreover, even if the injection scale of the isotropic driving is $\sim 100$~pc, the universal turbulent cascade can excite motions at the scale of individual molecular clouds \citep{Yuen22}. Within this study we do not address how turbulence anisotropy changes in multi-phase media \citep{instability}.
%https://ui.adsabs.harvard.edu/abs/2023MNRAS.521..230H/abstract
This and other questions will be addressed in subsequent studies.

\section{Summary}
\label{sec:sum}

 We explored how fundamental properties of incompressible MHD turbulence changed by the way of driving turbulence. We identified a new effect: the properties of turbulence radically change depending on whether it is driven through velocity or magnetic fluctuations. Our findings above can be briefly summarized as follows:
\begin{itemize}
     \item  The disparity of the energies of magnetic and velocity turbulence arises from the velocity driving of turbulence. The magnetic driving of turbulence provides similar amounts of energy in kinetic and magnetic fluctuations.
     \item  For velocity driving, fluid steams along stochastic magnetic field lines, and the kinetic energy of such a flow exceeds the energy of turbulent magnetic field fluctuations by a factor $\sim M_A^{-2}$.  Our comparison with earlier compressible MHD studies suggests that this effect arises from how turbulence is driven rather than fluid compressibility. We suggested a way of testing this hypothesis.
     \item For velocity driving of turbulence, fluid flow imprints the magnetic field stochasticity, resulting in the kinetic energy spectrum $E_K$ scaling $\sim k^{-2}$.
     The magnetic fluctuations slowly evolve near the injection scale corresponding to the $E_b\sim k^{-1}$ cascade. 
     \item Velocity-driven turbulence transfers more energy to slow modes than Alvenic modes, which induces numerous astrophysical implications. 
    \end{itemize}
The velocities and magnetic field spectra coincide if the turbulence is magnetically driven at the injection scale.

In terms of astrophysical implications, our study demonstrates
\begin{itemize}
  \item The importance of accounting of the nature of turbulence driving for evaluating the effects of turbulence for key astrophysical processes, e.g., on star formation, formation of filaments, cosmic ray propagation, and acceleration.
    \item  The traditional DCF approach for obtaining the magnetic field strength from observations must be modified depending on the turbulence driving. 
    \item We outline a possible way to determine the nature of turbulent driving from observations and show that measuring sonic and Alfven Mach numbers allows us to obtain the magnetic field using a universal expression that does not depend on the nature of turbulent driving. 
\end{itemize}

Our numerical testing is limited by incompressible MHD simulations. Some of the relations derived in this paper correspond to those obtained earlier empirically and were attributed to the effects of compressibility. {We hypothesize these effects can also arise due to velocity driving employed in the earlier studies. We predict that in the case of magnetic driving, the differences between kinetic and magnetic energies will be significantly reduced in the case of compressible simulations.
This calls for more studies exploring the extend that the fundamental properties of sub—Alfvenic MHD turbulence we explored in the incompressible regime carry over to the case when the compressibility is very important, i.e. for the case of turbulence in low-$\beta$ plasmas.}

\section*{Acknowledgements}
We thank Chris McKee for reading the paper and providing valuable suggestions on our presentation and discussions of the role of the slow modes in MHD turbulence.
We thank Andrey Beresnyak for discussing the effects of boundary conditions on the simulation results, Reinaldo Santos-Lima for discussing weak turbulence simulations,  and Dmitri Pogosyan for discussing the relation of parallel and perpendicular fluctuations. We also thank Jungyeon Cho and Hui Li for discussing $v_A$ variation and the nature of $k_\parallel=0$ modes in MHD turbulence. AL \& KWH acknowledges the support of NASA ATP AAH7546 and NSF 2307840. 
{ KHY thanks James Beattie for discussing the role of $k_\parallel= 0$ modes.} The research presented in this article was supported by the LDRD program of LANL with project \# 20220107DR (KWH) \& 20220700PRD1 (KHY), and a U.S. DOE Fusion Energy Science project. 
This research used resources from the LANL Institutional Computing Program (y23\_filaments), supported by the DOE NNSA Contract No. 89233218CNA000001. EV acknowledges the support from the AAS. 
This research also used resources of NERSC with award numbers FES-ERCAP-m4239 (PI: KHY) and m4364 (PI: KWH). We acknowledge Michael Halfmoon for additional NERSC time needed for this project. We also acknowledge the hospitality of KITP and its associated grant number NSF PHY-1748958.
\linebreak
{\noindent  Data Availability} The data underlying this article will be shared on reasonable request to the corresponding author.

\software{MHDFlows \citep{MHDFlows}, FourierFlows \citep{FourierFlows}}
\appendix 

\section{Properties of slow and fast modes}

 In \S \ref{2.2.1}, we discussed how fluctuations of velocities along magnetic field lines produce fluctuations in parallel magnetic field strength for incompressible turbulence, i.e., turbulence in the media corresponding to $\beta\rightarrow \infty$. The effect is generic for media with arbitrary $\beta$, however. 

\begin{figure*}
\centering
\centering
\includegraphics[width=0.80\linewidth]{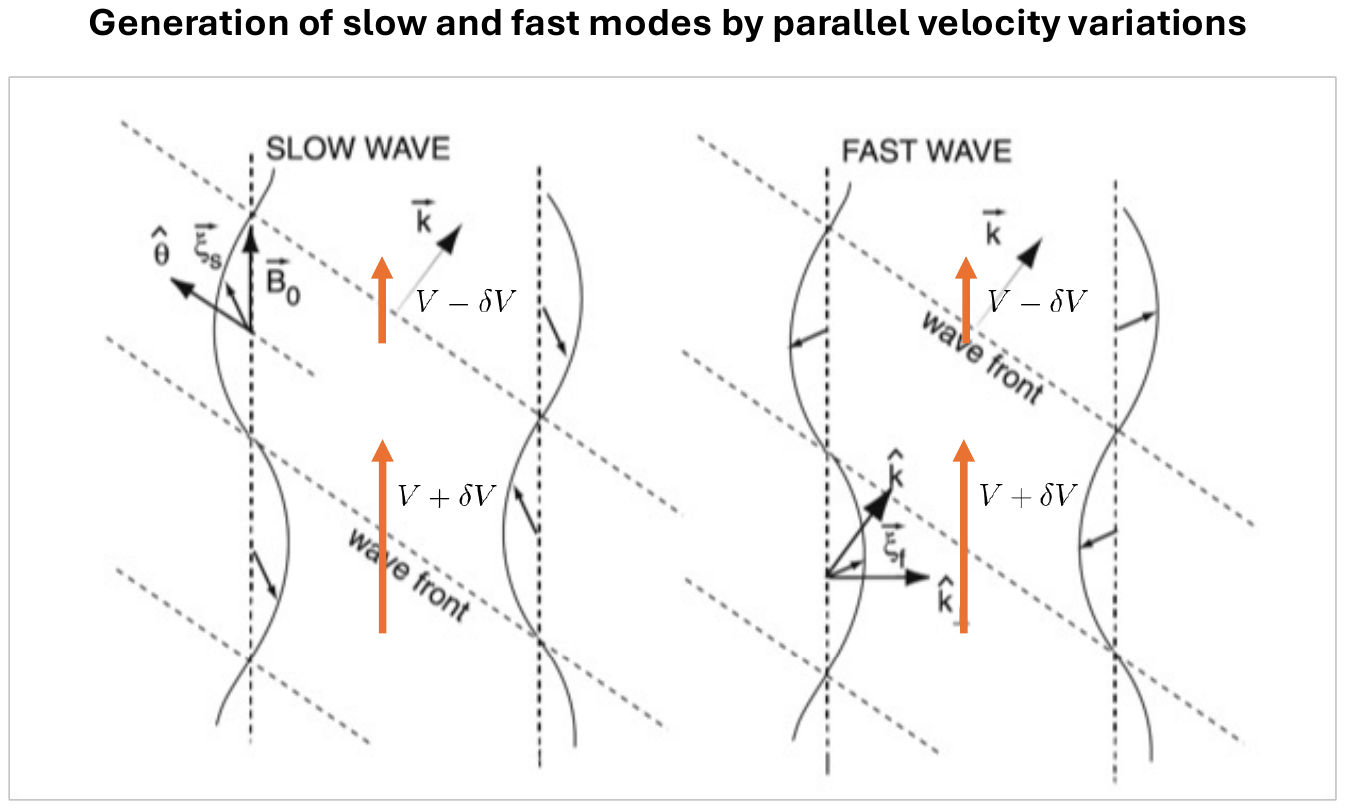}
\\ % Adds a line break between the two images
\caption{Red vectors correspond to the velocities of the flow along the flux tube. Slow and fast waves in real space (${ B}_0 - { k}$ plane). The directions of displacement vectors for a slow wave (left
panel) and a fast wave (right panel) are shown. Note that the transposition vector for the slow mode $\hat{\xi}_s$ lies between the wavefront $\hat{\theta}$ direction and ${ B}_0$
(= ${ k_\|}$) and $\hat{\xi_f}$ between ${ k}$ and ${ k_\bot}$. Note that $\hat{\theta}$ is perpendicular to ${ k}$ and parallel to the wavefront (modified from CLV03).}
\label{fig:slow/fast}
\end{figure*}

Figure \ref{fig:slow/fast} illustrates the effect. 
The mean magnetic field is in the vertical direction, and the fluid is pumped along the magnetic field. If the fluid flows along the magnetic field with mean velocity $V$, the variations of flow velocity $\delta V$ result in the variations of fluid pressure that are compensated by the compressions/expansions of the magnetic flux tube.
The corresponding variations of the flux tube thickness can be presented as a superposition of magnetosonic waves. 

Having a magnetic flux tube with thickness modulated by the flow, one can consider how slow and fast modes can be produced through such a modulation. Consider first a slow waves effect. Their wave vector ${ k}$ is perpendicular to the wavefront $\hat{\theta}$ and the vector of slow mode transposition $\vec{\zeta}_s$ 
is between $\hat{\theta}$ and ${ B}_0$. The angle between $\hat{\theta}$ and $\vec{\zeta}_s$ depends on plasma $\beta$, the two vectors get
parallel for $\beta\rightarrow \infty$ and get perpendicular for $\beta\rightarrow 0$. In the former case, slow modes are degenerate with Alfven waves, as they do not induce any compression of either fluid or magnetic field. 
The variations of the magnetic flux tube thickness shown in Figure \ref{fig:slow/fast} can be achieved by slow modes of different wavelengths and different directions of ${ k}$. This 
implies that the modulation of the flux tube thickness induced by the fluid (see Figure \ref{fig:slow/fast}) can generate a spectrum of slow waves with different ${ k}$. 
In particular, for the incompressible case, the excited slow modes with ${ k}$ perpendicular to ${ B}_0$ and correspond to the "condensate" with $k_\|=0$ that we study in our next paper. The exact composition and the properties of the magnetosonic waves that the Bernoulli effect can generate is beyond the scope of the present paper. 

The fast modes have their transposition vector $\vec{\zeta}_f$ between the direction ${ k}$ and the direction perpendicular to ${ B}_0$, i.e., the ${ k}_\bot$ direction. The relative orientation of ${ k}$ depends on plasma $\beta$. $\vec{\zeta}_f$ is aligned with ${ k}$ in incompressible fluids and get perpendicular to ${ k}$ when $\beta \rightarrow 0$. Fast modes do not exist for the incompressible fluid, as their speed is infinite. Thus, our paper does not consider the effect of fast waves. However, magnetosonic perturbations corresponding to fast modes become more important as $\beta$ decreases. 
For $\beta \rightarrow 0$ fast waves responsible for the modulation of magnetic flux as shown in Figure \ref{fig:slow/fast} are moving parallel to ${ B}_0$.

For a finite $\beta$, the fluctuation $\delta B_\|$ defined by Eq. (\ref{2}) is achieved by combining slow and fast modes with the composition of modes being a function of $\beta$. The contribution of fast modes is marginal for high $\beta$, but it gets dominant for $\beta \rightarrow 0$. In the latter case, slow modes correspond to sound waves propagating along magnetic field. Such waves cannot originate from the magnetic field compressions arising from the Bernoulli effect.

\bibliographystyle{aasjournal.bst}
\bibliography{xu}

\end{document}